\documentclass[journal, 10pt]{IEEEtran}
\usepackage[T1]{fontenc}
\usepackage[dvips]{graphicx}
\usepackage{relsize}
\usepackage{xcolor}
\usepackage{graphicx}
\usepackage{epsfig}
\usepackage{epstopdf}
\usepackage{amsmath}
\usepackage{amsfonts}
\usepackage{amssymb}
\usepackage{dsfont}
\usepackage{bbm}
\usepackage{mathtools,algorithm}
\usepackage{algpseudocode}
\usepackage{float} 
\usepackage{mathtools, cuted}
\usepackage{lipsum, color}
\usepackage{adjustbox}
\usepackage{multirow}
\usepackage{amssymb}
\usepackage{caption}
\usepackage[noadjust]{cite}

\DeclarePairedDelimiter\abs{\lvert}{\rvert}%
\begin{document}

	\title{Jitter Performance Evaluation for Resilient Vehicular Systems\vspace*{-0.3em}}
 \author{\IEEEauthorblockN{
 Pratiti Paul,~\IEEEmembership{Member,~IEEE,} Christo K. Thomas,~\IEEEmembership{Senior Member,~IEEE,} Walid Saad,~\IEEEmembership{Fellow,~IEEE,} Eric W. Burger,~\IEEEmembership{Fellow,~IEEE,} and Manav R. Bhatnagar,~\IEEEmembership{Senior Member,~IEEE}
 }
	\vspace*{-1.7em}
    \thanks{A preliminary version of this work \cite{asilomar} was presented at the IEEE Asilomar Conference on Signals, Systems, and Computers, October 2025. (Corresponding author: Pratiti
Paul.)

This work was supported by Virginia Tech Presidential Postdoctoral Fellowship Program, Commonwealth Cyber Initiative, and the Center for Assured and Resilient Navigation in Advanced TransportatION Systems (CARNATIONS) under the US Department of Transportation (USDOT)’s University Transportation Center (UTC) program (Grant No. 69A3552348324).

Pratiti Paul is with the Department of Electrical
	Engineering, Linköping University, Sweden. E-mail: pratiti.paul@liu.se.

Christo K. Thomas is with the Department of Electrical \& Computer Engineering, Worcester Polytechnic Institute, USA. E-mail: cthomas2@wpi.edu.

Walid Saad is with the Bradley Department of Electrical and Computer Engineering, Virginia Tech, USA. E-mail: walids@vt.edu.

Eric W. Burger is with the Commonwealth Cyber Initiative and the Bradley Department of Electrical and Computer Engineering, Virginia Tech, USA. E-mail: ewburger@vt.edu.

Manav R. Bhatnagar is
	with the Department of Electrical
	Engineering, Indian Institute of Technology (IIT) Delhi, India. E-mail:
	manav@ee.iitd.ac.in.
	}\vspace*{-1em}}
    
	\maketitle 

\begin{abstract}
In this paper, the challenge of resilient vehicular communications is investigated for a vehicle-to-vehicle (V2V) wireless link subject to timing jitter induced by the stochastic evolution of interference and inter-vehicular separation dynamics. To characterize the dynamic behavior of V2V systems under the influence of multiple stochastic deterioration processes, this work presents a novel mathematical framework for modeling and mitigating transmission delay jitter in V2V communication systems. A limit-state indicator is introduced to capture the progression of system performance, along with formal mathematical definitions of the system's jitter-withstanding capacity and the load-induced capacity degradation, defined in terms of the V2V system's ability to withstand jitter. These quantities provide the basis for tracking and assessing the system's resilience to timing jitter across all stages, including the alarming, failure, and restoration phases. To enhance resilience against jitter, adaptive power allocation and link diversity strategies (multiple-input-single-output) are investigated. These strategies bring the limit-state metric back within its safety bound, thereby resulting in an improved jitter-withstanding capacity. Numerical results validate the framework for quantifying jitter degradation and enabling resource-aware recovery. The results also establish that the investigated adaptive resource allocation scheme yields approximately 3-fold improvement in the average risk exposure rate relative to a constant resource allocation scheme baseline.
 \end{abstract}

\vspace{-0.3em}
\begin{IEEEkeywords}	
Chance constrained optimization, interference, jitter intolerance probability, jitter-withstanding capacity, limit-state indicator, mobility, resilience metrics, stochastic modeling, transmission delay jitter, vehicular communication systems.
\end{IEEEkeywords}	

\vspace{-1em}
\section{Introduction}
Vehicle-to-vehicle (V2V) communication is an essential enabler for next-generation intelligent transportation systems \cite{V2V}. V2V enables vehicles to wirelessly exchange real-time information such as speed, heading, position, and acceleration, facilitating safer, more efficient driving through early hazard detection and coordinated maneuvers. However, ensuring resilient V2V communication remains a significant challenge due to the inherently dynamic and mobile nature of vehicular environments. Under diverse traffic conditions, increases in both the mean inter-vehicular spacing or interference and its variability degrade communication reliability, which underscores the need for resilient control and communication mechanisms.

\vspace{-1em}
\subsection{Motivation} \vspace{-0.2em}
Real-world V2V communication operates under coupled spatial-channel dynamics that introduce significant uncertainty in packet transmission delays. On the one hand, V2V links, in practice, often function in regimes with large and dynamically varying inter-vehicular separations due to road and traffic conditions, visibility constraints, and behavioral variability among drivers. Such separations commonly occur in low-density highway or rural freeway traffic, on urban arterials where vehicles are spaced by signalized intersections or by sparse high-speed flows \cite{sparse}. Furthermore, IEEE 802.11p standard provides communication \cite{dist_jitter} at theoretical distances of up to $1000$ meters, both in V2V and vehicle-to-infrastructure (V2I) modes, and supports the frequency range of $5.85$--$5.925$ GHz. Whereas field experiments with IEEE $802.11$p \cite{dist_jitter} in a V2V scenario have shown that transmission delay jitter increases with inter-vehicular distance as propagation conditions deteriorate, and becomes more severe under mobility and channel-induced impairments such as fading and non-line-of-sight (NLoS) operation. Reported field results also show that when two vehicles communicated over an IEEE $802.11$p channel at $5.87$ GHz, jitter exceeds $100$ milliseconds at larger inter-vehicular separation distances for $500$-bytes data frames, which confirms that distance variation and adverse wireless channel conditions are critical contributors to timing instability in vehicular links. These observations reinforce the importance of studying jitter-resilient communication in safety-critical V2V systems. Additionally, different message types have distinct deadline requirements; for instance, steering and speed control messages typically require deadlines of $5$ milliseconds and $20$ milliseconds \cite{jitter_range}, respectively, while emergency braking and gear-shift messages are constrained by $40$ milliseconds and $20$ milliseconds deadlines \cite{jitter_range}, respectively. When vehicles move out of communication range, join or leave the network, or when inter-vehicular spacing becomes highly variable, the resulting jitter can exceed safe operational thresholds. On the other hand, the wireless medium becomes increasingly congested in dense traffic scenarios, where a high number of vehicles attempt to communicate over shared wireless channels. In the worst traffic conditions or during peak hours of the day, vehicles in close proximity often compete for the same spectral resources, which increases interference levels and results in fluctuations in signal-to-interference-plus-noise ratio (SINR) and packet transmission delay. The interplay between irregular inter-vehicular spacing and dynamic interference patterns produces varying transmission delay profiles, thus making V2V links more likely to suffer from intermittent connectivity disruptions.

Across all such conditions, the timely exchange of safety-critical messages becomes uncertain, thereby resulting in transmission delay variability, also known as \emph{transmission delay jitter}. Transmission delay jitter refers to variations in packet arrival times. In V2V systems, jitter can significantly disrupt the V2V link (for example, in emergency braking alerts \cite{ex3}, coordinated lane changes \cite{ex1}, intersection collision avoidance \cite{ex2}, real-time hazard warnings \cite{ex1}) by introducing random packet transmission delays that hinder timely responses. High jitter can delay safety beacons beyond their critical deadlines, while even low but inconsistent jitter can destabilize cooperative control systems like platooning and adaptive cruise control, which can delay critical safety messages, potentially leading to vehicle instability and increased accident risk. Therefore, all the preceding discussion motivates the development of resilient V2V communication systems capable of maintaining robust and sustainable transmission delay operation in the presence of unpredictable and adverse stressors.

\vspace{-.9em}
\subsection{Related work}
Several studies have examined the resilience of connected and autonomous vehicle (CAV) systems to cyber-physical disruptions, as well as the effects of communication and network jitter, typically as separate issues. In \cite{res1}, the authors used agent-based simulation to assess CAV performance during lane closures, communication failures, and signal outages, highlighting the role of V2V/V2I communication and redundancy in maintaining network efficiency. The work in \cite{res2} proposed a resilient event-triggered control framework for CAV platooning that compensates for communication delays and external disturbances and ensures robust leader-follower coordination. A microscopic simulation framework incorporating sensor degradation, control delays, and communication loss is introduced in \cite{res3} for qualifying time-dependent resilience using throughput and delay metrics. In \cite{res4}, the authors proposed a coupled vehicle infrastructure model to evaluate highway resilience. The work in \cite{res5} presented a multi-agent simulation that integrated adaptive cruise control, rerouting, and signal coordination to assess CAV robustness under diverse disruption types. A data-driven framework combining traffic flow models and trajectory data to capture both degradation and recovery is studied in \cite{res6}, emphasizing the importance of dynamic resilience indicators. In \cite{res7}, the authors employed a multi-layer simulation to assess CAV resilience to sensing and communication failures. The authors in \cite{res8} presented a macroscopic dynamic traffic assignment model, showing that CAV adoption enhances system-wide resilience. However, the existing literature \cite{res1}--\cite{res8} does not explicitly address the problem of timing jitter in resilient vehicular communication systems.
 
Meanwhile, jitter in existing studies is primarily addressed from higher layer networking or analytical perspectives rather than as a wireless link design challenge for V2V communication. The works in \cite{Chen2010, Kang2010, Torre2020, Zhang2019} treat jitter mainly as a queueing, buffering, and scheduling mechanism aimed at improving packet ordering and end-to-end delay performance. However, jitter due to physical layer wireless impairments, such as fading and mobility-induced channel variation, is not considered. Physical and link layer analyses of jitter and delay are presented in \cite{Gezici2007} and \cite{Wu2003}, which include timing jitter effects on error performance and delay bounds under fading. However, these studies remain analytical and do not develop jitter-resilient V2V transmission mechanisms. Similarly, in \cite{Szymanski2009, Park2012, Waqar2018, Chaporkar2007, Epiphaniou2010, Baz2015, Szymanski2009b}, the authors address the challenges of jitter through buffering, traffic shaping, statistical QoS provisioning, and E2E delay analysis in general or multi-hop networks, but they do not capture the mobility-driven and interference-sensitive delay fluctuations inherent to V2V links. The impact of delay and jitter on system level behavior, such as control stability, relay performance, packet delay trends, LTE load response, and low jitter traffic support, is investigated in \cite{Liu2021, Kandeepan2011, Ahmad2010, Mesbahi2016, Szymanski2007}, but it is not studied how configurable wireless transmission or link parameters may be leveraged to mitigate transmission delay jitter. Vehicular networking is considered in \cite{Li2016} and \cite{Liu2012} from a delay-tolerant routing perspective under intermittent connectivity, which targets delay tolerance rather than jitter-resilient V2V transmission. Overall, the existing studies do not explicitly investigate the impact of transmission delay jitter on V2V wireless links under mobility and interference uncertainty. Moreover, whether configurable communication resources can be adjusted to reduce jitter and enhance link resilience remains largely unexplored.

 In our previous work \cite{asilomar}, we analyzed the impact of V2V variations on jitter and introduced a mathematical framework in which a key performance indicator (KPI) quantifies jitter resilience. However, our prior work in \cite{asilomar} does not account for the effect of random interference in the jitter characterization. It also does not consider the analysis of the adaptive allocation mechanism of optimized resources in the post-failure phase to enhance the jitter performance of the V2V system.

\vspace{-1.em}
\subsection{Contribution}
The main contribution of this paper is the development of a novel performance metric for jitter-resilient V2V communication systems. To address the challenge of transmission delay jitter in V2V systems, we introduce a probabilistic measure called jitter intolerance probability for quantifying the system's ability to tolerate jitter induced by external stressors. Building on this, we analytically formalize the limit-state indicator function to serve as a benchmark for determining whether system performance remains above a defined operational threshold. The modeling begins by representing transmission delay jitter as a state-dependent stochastic process driven by two key factors: increasing inter-vehicular distance-dependent fluctuations and rising interference from surrounding vehicles. These dynamics are shown to contribute to jitter intolerance performance deterioration, which we capture through a risk-based analysis using newly formulated resilience metrics. This established framework enables a comprehensive analysis of the full resilience cycle, spanning normal operation, performance degradation, adaptive response, and recovery, modeled using a non-homogeneous Markov process with state-dependent transitions. The system's stochastic evolution is characterized using a set of performance indicators, including the limit-state indicator function, jitter-withstanding capacity, and the jitter-induced load. The jitter-withstanding capacity is defined as the level of jitter variability the V2V system can tolerate while maintaining reliable and timely data exchange for safe vehicle coordination. Simulation results evaluate system behavior across different resilience phases and show how performance can be significantly improved through timely adaptation, adaptive power allocation, and the use of diversity-enhancing strategies such as multiple-input single-output (MISO) configurations. The proposed resource allocation framework is developed using a chance-constrained optimization approach, and adaptive resource allocation is shown to provide superior jitter tolerance compared with constant allocation by dynamically responding to stressor variability.
\begin{figure}[t]
\centerline{\psfig{file=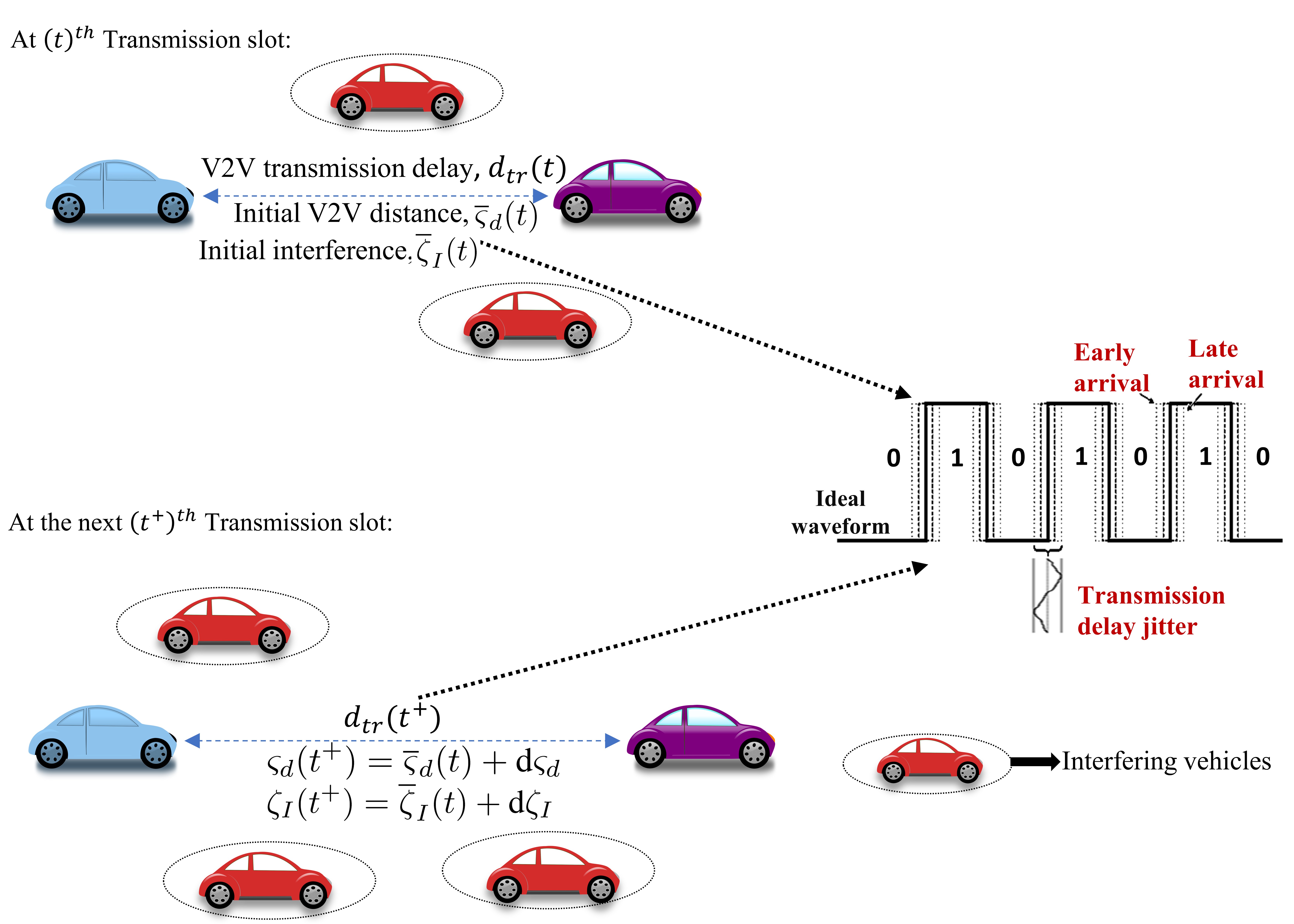,width=3.4 in, height = 2.4 in }}
    \vspace{-0.3em}
		\caption{\small V2V system affected by transmission delay jitter from random interference and varying inter-vehicular distances.}
		\vspace{-1.7em}
		\label{sys_mod}
\end{figure} 

The rest of the paper is organized as follows. Section II introduces the stochastic V2V system model and presents the mathematical formulation of the deterioration processes and their impact on the jitter performance. Section III presents the analytical framework of the system's jitter state and its stochastic evolution in the presence of stressors. A risk-based analysis of the system state and formulation of the limit-state are provided in Section IV, while Section V develops a resource allocation framework for jitter-resilient V2V communication. Numerical results and concluding remarks are presented in Sections VI and VII, respectively.

\vspace{-0.6em}
\section{System Model}
Consider a single V2V wireless link that experiences long-term, persistently increasing transmission delay variations or jitter. As illustrated in Fig.~\ref{sys_mod}, jitter stems from two main factors: (a) the dynamically varying distance between the two vehicles, and (b) the random appearance of interfering vehicles over time. Therefore, the stochastic variations of the aforementioned deterioration events consistently intensify the temporal fluctuations in packet transmission delay over the wireless communication link. As a result, data packets arrive at the destination vehicle outside their expected time windows, which in turn degrades the communication reliability. To quantify jitter performance in the V2V communication context, we develop a stochastic framework along with a theoretically tractable metric that captures its temporal evolution.

To compute transmission jitter, we define a discrete-time system state indexed by the transmission slot, assuming that packets are sent in a slotted manner. The corresponding instantaneous SINR of the V2V link in slot $t$ will be:\vspace{-0.5em}
\begin{IEEEeqnarray}{rCl}
\gamma(t)=\frac{P h^2}{N_0+ \zeta_{I}(t)} \bigg(\frac{\lambda_{\textrm{op}}}{4\pi \varsigma_d(t)}\bigg)^2,\label{SINR} \vspace{-0.5em}
\end{IEEEeqnarray}
where $\varsigma_d(t)$ is the random inter-vehicular distance between two vehicles and $h$ is the fading channel gain of the considered V2V link. In the considered V2V setup, we assume NLoS propagation with rich scattering at both the transmitter and receiver, resulting in a Rayleigh-distributed 
channel \cite{Ray_scatter_rich, Ray_scatter_rich1, Ray_scatter_rich2} gain with unit mean power, i.e., $\mathbb{E}[\abs{h}^2]=1$. This 
assumption is typical for urban V2V scenarios with building 
obstructions. Empirical measurements on highway LoS links at $5.9$ GHz show that 
small-scale fading is better characterized by Rician distributions 
(with Rice factor $\kappa$ $\geq 10$ dB) rather than Rayleigh \cite{rodrigo2016analysis,sen2008vehicle}. Thus, the Rayleigh assumption represents a conservative
degradation estimate for highway scenarios, as it ignores the beneficial dominant LoS component. Our framework can be extended to Rician fading by modifying the channel gain statistics in \eqref{SINR}, but the degradation analysis in the next sections remains qualitatively similar. $\lambda_{\textrm{op}}$ represents the operating spectrum band, $P$ is the transmit power, and $N_0$ is the noise power spectral density for additive Gaussian noise at the receiver of the destination vehicle. The interference of the considered link at time instant $t$ can be expressed as $\zeta_{I}(t)=\sum_{\substack{j=1}}^{N(t)}P_j h_j^2 (d_j)^{-\alpha}$, where $P_j$, $h_j$, and $d_j$ represent the transmit power, fading channel gain, and the distance of the $j^{\mathrm{th}}$ interfering vehicle from the destination vehicle, respectively. $\alpha$ is the path loss coefficient. $N(t)$ is the number of random interfering vehicles generating interference with the active V2V link at time instant $t$. The random transmission delay for a packet transmission at time instant $t$ is $	
d_{\mathrm{tr}}(t)=\frac{L_p}{B \log_2(1+\gamma(t))}$, where $L_p$ is the transmission packet size and $B$ is the bandwidth. Transmission jitter is therefore quantified as the variation in transmission delay between successive transmission slots as: \vspace{-0.7em}
\begin{IEEEeqnarray}{rCl}	
\tau_{\mathrm{tr}}(t^+) = \big(d_{\mathrm{tr}}(t^+)-d_{\mathrm{tr}}(t)\big),\label{jit} \vspace{-0.7em}
\end{IEEEeqnarray}
where $t$ and $t^+$ denote the immediately preceding and subsequent transmission slots, respectively. Additionally, $\tau_{\mathrm{tr}}$ is a random variable whose temporal fluctuations give rise to jitter. Increasing variability in inter-vehicular distance and interference exacerbates uncertainty in packet timing, which leads to elevated jitter over the V2V link. 
\vspace{-1.5em}
\subsection{Deterioration Growth Modeling}
In the considered system, interference dynamics are modeled via a Poisson arrival process of interfering vehicles \cite{PP_VT}, while inter-vehicular incremental distance variation follows a diffusion process \cite{drift_diffusion}.

\subsubsection{Interference growth modeling} Let the initial mean arrival rate of the interferer be $\lambda_0$. During rush-hour traffic in urban environments, peak channel demand and network congestion significantly intensify interference on dedicated V2V communication links \cite{congested_interference}. Therefore, since the objective is to quantify the growing impact of interference, we adopt a pure-birth population model \cite{birth} in which vehicles enter the interference zone, but departures are neglected within the critical time horizon considered. This assumption highlights the cumulative effect of interference and yields a conservative estimate of performance degradation. For a considered interfering vehicle $j$, the instantaneous power per interferer is $\nu_{j} =
P_j h_{j}^2 d_{j}^{-\alpha}$. Also, the total number of interfering vehicles appeared at the $t^+$ time instant is $
m_{\delta} = I_{0}(t) + \lambda(t),$ where $I_0$ is the initial interferer count following the initial Poisson rate $\lambda_0$ at previous time instant $t$. $\lambda(t)=\omega \lambda_0 t^\varphi$ is the additional interfering vehicles that appear during the transition from $t$ to $t^+$ instant, and this value is assumed to grow with time. Hence, the aggregate interference power at time instant $t^+$ is given as:\vspace{-0.9em}
\begin{IEEEeqnarray}{rCl}
\hspace{-1em}\zeta_I(t^+) &=& \sum\limits_{j=1}^{ m_{\delta}}\nu_{j}= \sum\limits_{j=1}^{I_0(t)}P_j h_{j}^2 d_{j}^{-\alpha}+ \sum\limits_{j=1}^{\lambda(t)}P_j h_{j}^2 d_{j}^{-\alpha}. \label{mean_int_increment} \vspace{-0.7em}
\end{IEEEeqnarray}
$\zeta_I$ is a stochastic interference evolution process driven by the same underlying Poisson arrival process with random increments. The successive growth of interference between $t$ and $t^+$ time interval due to growing congestion can be obtained as $\text{d}\zeta_I = \zeta_I(t^+)-\overline{\zeta}_I(t)$, where $\overline{\zeta}_I(t)\bigg(=\sum\limits_{j=1}^{I_0(t)}P_j h_{j}^2 d_{j}^{-\alpha}\bigg)$ is the total instantaneous interference power in the previous time slot $t$ for the $I_0$ interfering vehicles.

\vspace{0.5em}
\subsubsection{Inter-vehicular distance growth modeling} We next model a second key deterioration mechanism, which is the variation in the distance between the two vehicles. Specifically, we treat this inter-vehicular spacing as a random degradation process that directly drives delay variation by capturing both its mean drift and its stochastic fluctuations. To characterize how the variance of those fluctuations themselves changes with traffic conditions, we adopt an Itô drift-diffusion model \cite{ito}. Let $\varsigma_d(t)$ be the instantaneous distance between two vehicles at time $t\ge0$. Let $W_t$ and $\Theta_t$ be independent standard Brownian motions (Wiener processes \cite{GBM}) whose increments are Gaussian with zero mean and variance $\text{d}t$. Here, $\Theta_t$ drives the instantaneous distance fluctuations, whereas $W_t$ governs the evolution of their variance. We model the distance $\varsigma_d$ (where initial distance at $t$ time instant is $d_0$) and variance driving the small-time fluctuations of $\varsigma_d$ as $v_t$, with two coupled stochastic differential equations (SDE) as:\vspace{-0.6em}
\begin{IEEEeqnarray}{rCl} \vspace{-0.7em}
\text{d}\varsigma_d(t) &=& \mu(t) \text{d}t + \sqrt{v(t)}\text{d}|\Theta_t|, 
\label{eq_dv2v}
\end{IEEEeqnarray}
\begin{IEEEeqnarray}{rCl}
\text{d}v(t) &=& \eta v(t) \text{d}t + \xi v(t) \text{d}W_t, \label{dist_fluc} \vspace{-0.5em}
\end{IEEEeqnarray}
where $\mu(t)$ is the mean, initial variance level $v_0>0$, variance growth‑rate $\eta>0$, and volatility of variance $\xi>0$. \textcolor{black}{Here, the parameter $\eta$ represents the rate at which the variance of inter-vehicular distance fluctuations grows, effectively capturing how quickly uncertainty increases with time. The parameter $\xi$ reflects the volatility or intensity of these fluctuations, indicating how widely the instantaneous inter-vehicular spacing may vary around its average trend.} $|\Theta_t|$ signifies the increment of mean V2V distance with time. It is considered that $\varsigma_d$ is strictly positive, since a negative distance has no physical significance in the considered scenario. Therefore, integrating over \([t,t^+]\), we get 
$\varsigma_d(t^+)
 = d_0+\int_{t}^{t^+}\mu(t) \text{d}t
 +\int_{t}^{t^+}\sqrt{v(t)} \text{d}|\Theta_t|$.
The deterministic mean trajectory is then
defined by the mean path traversed as: \vspace{-1.2em}
\begin{IEEEeqnarray}{rCl}
\overline{\varsigma}_d(t)
 =\mathbb{E}[\varsigma_d(t)]
 = d_0+\int_{t}^{t^+}\mu(t) \text{d}t . \label{dist_bar} \vspace{-0.5em}
\end{IEEEeqnarray}
And for the variance process $v(t)$ (a geometric Brownian motion) we have  
$\mathbb{E}[v(t)]=v_0e^{\eta t}$. Hence:\vspace{-0.5em}
\begin{IEEEeqnarray}{rCl}
\sigma_{d}^{2}(t)
 =\int_{t}^{t^+}\mathbb{E}[v_s] \text{d}s
   =\frac{v_0}{\eta}\bigl(e^{\eta t}-1\bigr). \label{dist_GBM}\vspace{-0.5em}
\end{IEEEeqnarray}
We represent the stochastic zero‑mean fluctuation in \eqref{eq_dv2v} by:
$\Theta_t^{*}=\bigl[\int_{t}^{t^+}\sqrt{v(t)} \text{d}\Theta_t\bigr]/\sigma_{d}(t),$ so that $\Theta_t^{*}\sim\mathcal{N}(0,1)$. Thus we have
$\int_{t}^{t^+}\sqrt{v(t)} \text{d}|\Theta_t|
   =\sigma_{d}(t) |\Theta_t^{*}|$. Since $\Theta_t^{*}$ follows a normal distribution, $|\Theta_t^{*}|$ follows a half-normal distribution.  Thereby, the inter-vehicular distance evolved with explicit mean and variance at time instant $t^+$ can be obtained as:\vspace{-0.7em}
\begin{IEEEeqnarray}{rCl}
\hspace{-1.7em}\varsigma_d(t^+)
=\Bigl[d_0+\int_{t}^{t^+}\mu_s\,\text{d}s\Bigr]
+\sqrt{\frac{v_0}{\eta}\bigl(e^{\eta t}-1\bigr)} |\Theta_t^{*}|. \label{dist_evol} \vspace{-0.5em}
\end{IEEEeqnarray}
$\varsigma_d$ is a stochastic process for random V2V distance evolution. This expression models $\varsigma_d(t^+)$ as the sum of its deterministic mean trajectory $\overline{\varsigma}_d(t)$ and a zero‑mean Gaussian fluctuation whose standard deviation is $\sigma_{d}(t)$. Variation in distance across consecutive time slots can be formulated as $\text{d}\varsigma_d = \varsigma_d(t^+)-\overline{\varsigma}_d(t)$. $\overline{\varsigma}_d(t)$ is considered to be the distance at the previous time instant. The considered stressors: interference intensity and V2V distance variations are non-decreasing functions of time for the considered V2V system during the congestion-enduring time. We next model the dependency of jitter as a function of the evolving mobility-induced distance fluctuations and network-induced interference variations.  

\vspace{-1.em}
\subsection{Jitter State Evolution with Deterioration}
\textcolor{black}{The jitter state evolves incrementally over time, with each transition driven by the stochastic growth of interference and inter-vehicular distance. Thus, the magnitude of each jitter increment depends on the system's current state, thereby making the evolution state-dependent. Furthermore, since the intensities of both stressors grow with time, the transition probabilities are non-stationary. These characteristics motivate modeling the jitter state evolution as a non-homogeneous state-dependent Markov process \cite{paolo}, as captured below:\vspace{-0.5em}}
\begin{IEEEeqnarray}{rCl}	
\tau_{\mathrm{tr}}(t^+)&=& \tau_{\mathrm{tr}}(t) + p_I p_d \mathcal{S}[\tau_{\mathrm{tr}}(t)] (\text{d} \zeta_I + \text{d} \varsigma_d) +\hspace{-0.3em} (1-p_I) p_d  \nonumber\\&\times&\mathcal{S'}[\tau_{\mathrm{tr}}(t)] \text{d} \varsigma_d+\hspace{-0.3em}(1-p_d) p_I  \mathcal{S''}[\tau_{\mathrm{tr}}(t)] \text{d} \zeta_{I},
\label{next_jit} \vspace{-0.3em}
\end{IEEEeqnarray}
where the functions $\mathcal{S}[\cdot]$, $\mathcal{S'}[\cdot]$, and $\mathcal{S''}[\cdot]$ are the non-negative, finite-valued mappings of the system's jitter state, dependent on the current jitter state $\tau_{\mathrm{tr}}(t)$ and the instantaneous impacts of $\zeta_I$ and $\varsigma_d$ during the transition period from $t$ to $t^+$. $p_d$ and $p_I$ are the probability of the occurrence of jitter state change due to distance fluctuations and interference, respectively, i.e., the states where $\varsigma_d(t^+)\neq \varsigma_d(t)$. Additionally, the fourth event, $(1-p_I)(1-p_d)$, is not considered, as it corresponds to a scenario in which the deterioration processes do not impact the jitter state and, thereby, the system's jitter state does not vary from its previous state. Thus, the resulting jitter states under different disturbance variations can be explicitly expressed as:

\vspace{-1.7em}
{\small \begin{IEEEeqnarray}{rCl}
&\mathcal{S}&[\tau_{\mathrm{tr}}(t)] (\text{d} \zeta_I +\text{d} \varsigma_d)
=  \tau_{\mathrm{tr}}(t^+) - \tau_{\mathrm{tr}}(t) =\int_{t}^{t^+} \text{d}\tau_{\mathrm{tr}}(t)
 \nonumber\\&=& 
\underbrace{\int_{t}^{t^+}\bigg[\frac{\partial \tau_{\mathrm{tr}}(t)}{\partial \zeta_I}\,\text{d}\zeta_I
  +\frac{\partial \tau_{\mathrm{tr}}(t)}{\partial \varsigma_d}\,\text{d}\varsigma_d\bigg]}_{\text{Joint stochastic effect of distance \& interference variation}}. \label{state_dist_int}
\end{IEEEeqnarray}}

\noindent And also, 
\vspace{-1.5em}
{\small \begin{IEEEeqnarray}{rCl}
\mathcal{S'}[\tau_{\mathrm{tr}}(t)] \text{d} \varsigma_d
&=& \hspace{-1em}
\underbrace{\int_{t}^{t^+}\bigg[\frac{\partial \tau_{\mathrm{tr}}(t)}{\partial \varsigma_d}\,\text{d}\varsigma_d
  \bigg]}_{\text{Stochastic effect of distance variation only}},~~~~~\label{state_dist} \vspace{-1.4em}
\end{IEEEeqnarray}
\begin{IEEEeqnarray}{rCl}
\mathcal{S''}[\tau_{\mathrm{tr}}(t)] \text{d} \zeta_{I}
&=& \hspace{-1em}
  \underbrace{\int_{t}^{t^+}\bigg[\frac{\partial \tau_{\mathrm{tr}}(t)}{\partial \zeta_{I}}\,\text{d}\zeta_{I}
  \bigg]}_{\text{Stochastic effect of interference variation only}}.~~~~~ \label{state_interfer}
\end{IEEEeqnarray}}

\noindent In practice, changes in inter-vehicular distance and the arrival of new interfering vehicles occur on a time scale that is orders of magnitude longer than a single packet transmission slot. Consequently, each degraded state must be characterized by its random sojourn time, which typically spans many transmission slots. Let $\{\tau_{\mathrm{tr}}(t)\}_{t{\ge0}}$ be the discrete-time jitter state process, and assume a new persistent jitter state $\mathcal{S}_p$ emerges at a given time instant due to one or both of the deterioration mechanisms. Hence, $T_p(t)$ is the random sojourn time of the newly entered jitter state, i.e., the number of consecutive time slots the system stays in that jitter state, and can further be formally expressed as:\vspace{-0.7em}
\begin{IEEEeqnarray}{rCl}
T_p(t) = \min\bigl\{m\ge 1 : \tau_{\mathrm{tr}}(t+m)\neq \mathcal{S}_p,\bigr\}. \vspace{-0.7em}
\end{IEEEeqnarray}
The sojourn time is exponentially distributed with rate parameter 
$p_H$ (normalized: $\sum\limits_{n=1}^\infty \Pr(T_p = n) = 1)$. This geometric distribution 
arises naturally when state transitions occur according to a 
Poisson-modulated jump process on the underlying vehicle positions 
and interference patterns. The mean sojourn time is:
$\mathbb{E}[T_p] = p_H^{-1}$     [in units of time slots].
This modeling choice implies that state transitions are memoryless, 
a reasonable approximation when vehicle arrival/departure and interference dynamics are well-modeled by renewal processes. Thus, $\tau_{\mathrm{tr}}(t)$ remains in state $\mathcal{S}_p$ for the interval $t,\dots,t+n-1$ and transitions to a different state at $t+n$. The transition probability from the current jitter state at $t$ to the next state at $(t + t^+)$ is characterized by the corresponding jitter intolerance probability $\mathcal{P}_{JI}$, which reflects the likelihood of state change under the influence of deterioration processes. Therefore, $\mathcal{P}_{JI}$ being a function of jitter state evolving from the previous state to a more degraded next state can be formulated for each state change at a particular time instant as:\vspace{-0.7em}
\begin{IEEEeqnarray}{rCl}	
&&\hspace{-1.1em}\mathcal{P}_{JI}^+(t) = 
p_I p_d \Pr\bigl[\mathcal{S}[\tau_{\mathrm{tr}}(t)]| \text{d}\varsigma_d> 0,\text{d}\zeta_{I}>0\bigr] \nonumber\\&\times&\Pr\bigl[\text{d}\varsigma_d>0\bigl] \Pr\bigl[\text{d}\zeta_{I}>0\bigl] + (1-p_I)p_d \nonumber\\&\times&\Pr\bigl[\mathcal{S'}[\tau_{\mathrm{tr}}(t)]|\text{d}\varsigma_d>0\bigr] \Pr\bigl[ \text{d}\varsigma_d>0\bigl]  \nonumber\\&+& (1-p_d)p_I \Pr\bigl[\mathcal{S''}[\tau_{\mathrm{tr}}(t)]|\text{d}\zeta_{I}>0\bigr] \Pr\bigl[\text{d}\zeta_{I}>0\bigl],
\label{jit_intol} \vspace{-0.5em}
\end{IEEEeqnarray}
where $(\cdot)^+$ denotes the transition from $t$ to $t^+$ time instant.

To make the considered V2V communication system resilient to jitter, we structure the problem into three main steps. First (Section III), we probabilistically derive the stressor dynamics and characterize the resulting statistical distribution of the jitter state evolution. This step establishes an understanding of how various deterioration processes impact the jitter performance of the V2V link and how such degradation can potentially push the system toward failure. Second (Section IV), we introduce a key performance indicator-based resilience metric that explicitly distinguishes between different regions of the resilience cycle, namely alarming, failure, and restoration. This indicator function intuitively captures the communication sustainability of the V2V link amid several random deterioration processes. Finally (Section V), we address restoration of system performance by minimizing jitter through optimized allocation of communication resources.

\vspace{-0.7em}
\section{V2V Transmission Delay Jitter Analysis}    
In this section, the underlying deterioration processes and their statistical distributions are first analyzed for the considered V2V system. Subsequently, the impact of these evolving deterioration processes on the transition of the jitter state is analytically derived. It is assumed that $\varsigma_d$ (from \eqref{dist_evol}) is the random increment of the V2V distance. So, $\varsigma_d$ takes a positive value around its mean distance $\overline{\varsigma}_d(t^-)$ at $t^{\mathrm{th}}$ instant, and the V2V distances are correlated for every two subsequent transmission slots. Also, from \eqref{dist_evol}, it can be observed that $\varsigma_d$ follows a shifted half-normal distribution. Therefore, the probability distribution function (PDF) of the stepwise increment of V2V distance variation is computed from \eqref{dist_bar} and \eqref{dist_GBM} as:\vspace{-0.8em} 
\begin{IEEEeqnarray}{rCl}	
f_{\varsigma_d}(d;t) &=& \frac{\sqrt{2}}{\big|\sqrt{\frac{v_0}{\eta}\big(e^{\eta t}-1\big)}\big|\sqrt{\pi}} e^{\bigg(-\frac{(d-\overline{\varsigma}_d(t))^2}{\frac{2 v_0}{\eta}\big(e^{\eta t}-1\big)}\bigg)},~~~d>0.~~~~~\label{dist_pdf} \vspace{-0.6em}
\end{IEEEeqnarray}
Therefore, the stochastic mean and variance functions of the deterioration processes can be mathematically expressed as follows:\vspace{-0.9em}
\begin{IEEEeqnarray}{rCl}	
\mu_{\varsigma_d}(t)=\mathbb{E}[\varsigma_d] = \overline{\varsigma}_d(t) + \sqrt{\frac{2v_0}{\pi \eta}(e^{\eta t}-1)}. \label{mu_dist} \vspace{-1em}
\end{IEEEeqnarray}
\begin{IEEEeqnarray}{rCl}	
\sigma_{\varsigma_d}^2(t)=\mathbb{V}[\varsigma_d] = \frac{v_0}{\eta}(e^{\eta t}-1)\bigg(1-\frac{2}{\pi}\bigg).\label{sigma_dist} \vspace{-0.5em}
\end{IEEEeqnarray}
Also, for evolving interference $\zeta_I=\sum\limits_{j=1}^{I_{0}(t) + \lambda(t)}\Bigl(P_j h_{j}^2 d_{j}^{-\alpha}\Bigr)$, which is a summation of square of independent and identically distributed (i.i.d.) Rayleigh random variables with corresponding weights. Therefore, the PDF of the interference growth from \eqref{mean_int_increment} is derived as follows:  \vspace{-0.7em}
\begin{IEEEeqnarray}{rCl}	f_{\zeta_{I}}(l;t) &=& \sum_{j=1}^{I_{0}(t) + \lambda(t)} \frac{\prod_{j=1}^{I_{0}(t) + \lambda(t)} (1/2\beta_j)}{\prod_{\substack{k=1 \\ k\neq j}}^{I_{0}(t) + \lambda(t)}(1/2\beta_k-1/2\beta_j)} \nonumber\\&\times&\exp{\bigg(\frac{-l}{2\beta_j}\bigg)}, \label{int_pdf}
\end{IEEEeqnarray}
where $\beta_j=P_j (d_j)^{-a}$. Next, considering $w_j = \frac{\prod_{j=1}^{I_{0}(t) + \lambda(t)} (1/2\beta_j)}{\prod_{\substack{k=1 \\ k\neq j}}^{I_{0}(t) + \lambda(t)}(1/2\beta_k-1/2\beta_j)}$, we have: \vspace{-1.4em}
\begin{IEEEeqnarray}{rCl}	
\mu_{\zeta_I}(t)=\mathbb{E}[\zeta_I] = \sum_{j=1}^{I_{0}(t) + \lambda(t)} 2 w_j \beta_j. \label{mu_int} \vspace{-1em}
\end{IEEEeqnarray}\vspace{-0.3em}
\begin{IEEEeqnarray}{rCl}	
\hspace{-1.5em}\sigma_{\zeta_I}^2(t)=\mathbb{V}[\zeta_I] = \sum_{j=1}^{I_{0}(t) + \lambda(t)} \hspace{-1em}8 w_j \beta_j^2 - \bigg(\sum_{j=1}^{I_{0}(t) + \lambda(t)} \hspace{-1em} 2 w_j \beta_j\bigg)^2.~~~\label{sigma_int} \vspace{-0.7em}
\end{IEEEeqnarray}
Thus, $\varsigma_d$ and $\zeta_I$ capture the statistical behavior of V2V distance variation and interference evolution, respectively, during each transition from the previous jitter state to the next jitter state of the considered V2V system. We next derive the distribution 
of the resulting jitter $\tau_{\mathrm{tr}}$. Since $d_{\mathrm{tr}}(t)$ is a nonlinear function of SINR, thereby implementing the first-order Taylor expansion (Delta method), it can further be expressed as:\vspace{-0.5em}
\begin{IEEEeqnarray}{rCl}	
   d_{\mathrm{tr}}(t) \approx d_{\mathrm{tr}}(t^{-}) 
+ \frac{\partial d_{\mathrm{tr}}}{\partial \zeta_I}\,\mathrm{d}\zeta_I 
+ \frac{\partial d_{\mathrm{tr}}}{\partial \varsigma_d}\,\mathrm{d}\varsigma_d , \vspace{-0.5em}
\end{IEEEeqnarray}
where $d_{\mathrm{tr}}(t^{-})$ is the delay in the previous slot and $\mathrm{d}\zeta_I$, $\mathrm{d}\varsigma_d$ are the increments in interference and distance variations. The jitter is then the change in delay between successive slots, which makes it a function of these stressor increments. We can further illustrate \eqref{jit} as the transmission delay jitter between $t$ and $t^+$ of the considered V2V link caused by random interference and V2V distance variations, and can be expressed in \eqref{next_jit_exp}, as shown at the top of the next page. From \eqref{next_jit}, the jitter state for all possible combinations of deterioration events can be explicitly written in a more generalized form as shown at the top of the next page.
\begin{figure*}[t!] \vspace{-1.5em}
\smaller{\begin{IEEEeqnarray}{rCl}	
&&\hspace{-0.7em}\tau_{\mathrm{tr}}(t) \hspace{-0.3em}=\hspace{-0.3em} \frac{p_I p_d L_p}{B} \hspace{-0.26em}\Bigg[\hspace{-0.15em}\frac{1}{ \log_2\hspace{-0.3em}\bigg(\hspace{-0.3em}1\hspace{-0.3em}+\hspace{-0.3em}\bigg(\hspace{-0.3em}\frac{P h^2}{N_0+ \zeta_I(t)}\hspace{-0.3em} \bigg(\hspace{-0.3em}\frac{\lambda_{\textrm{op}}}{4\pi \varsigma_d(t)}\hspace{-0.3em}\bigg)^2\bigg)\hspace{-0.3em}\bigg)} \hspace{-0.3em}-\hspace{-0.3em} \frac{1}{ \log_2\hspace{-0.3em}\bigg(\hspace{-0.3em}1\hspace{-0.3em}+\hspace{-0.3em}\bigg(\hspace{-0.3em}\frac{P h^2}{N_0+ \overline{\zeta}_{I}(t^-)} \bigg(\hspace{-0.3em}\frac{\lambda_{\textrm{op}}}{4\pi \overline{\varsigma}_d(t^-)}\hspace{-0.3em}\bigg)^2\bigg)\hspace{-0.3em}\bigg)}\hspace{-0.3em}\Bigg] \hspace{-0.3em}+\hspace{-0.3em} \frac{p_I (1\hspace{-0.3em}-\hspace{-0.3em}p_d) L_p}{B}\hspace{-0.3em}\Bigg[\hspace{-0.1em}\frac{1}{\log_2\hspace{-0.3em}\bigg(\hspace{-0.3em}1\hspace{-0.3em}+\hspace{-0.3em}\bigg(\frac{P h^2}{N_0 + \zeta_I(t)} \bigg(\frac{\lambda_{\textrm{op}}}{4\pi \overline{\varsigma}_d(t^-)}\bigg)^2\bigg)\hspace{-0.3em}\bigg)} \nonumber\\&-& \hspace{-0.3em}\frac{1}{ \log_2\hspace{-0.3em}\bigg(\hspace{-0.3em}1\hspace{-0.3em}+\hspace{-0.3em}\bigg(\hspace{-0.3em}\frac{P h^2}{N_0+ \overline{\zeta}_{I}(t^-)} \bigg(\hspace{-0.3em}\frac{\lambda_{\textrm{op}}}{4\pi \overline{\varsigma}_d(t^-)}\hspace{-0.3em}\bigg)^2\bigg)\hspace{-0.3em}\bigg)}\Bigg] \hspace{-0.3em}+\hspace{-0.3em} \frac{(1-p_I) p_d L_p}{B} \Bigg[\frac{1}{ \log_2\hspace{-0.3em}\bigg(\hspace{-0.3em}1\hspace{-0.3em}+\hspace{-0.3em}\bigg(\hspace{-0.3em}\frac{P h^2}{N_0 + \overline{\zeta}_{I}(t^-)} \bigg(\hspace{-0.3em}\frac{\lambda_{\textrm{op}}}{4\pi \varsigma_d(t)}\hspace{-0.3em}\bigg)^2\bigg)\hspace{-0.3em}\bigg)} \hspace{-0.3em}-\hspace{-0.3em} \frac{1}{ \log_2\hspace{-0.3em}\bigg(\hspace{-0.3em}1\hspace{-0.3em}+\hspace{-0.3em}\bigg(\frac{P h^2}{N_0 + \overline{\zeta}_{I}(t^-)} \bigg(\hspace{-0.3em}\frac{\lambda_{\textrm{op}}}{4\pi \overline{\varsigma}_d(t^-)}\hspace{-0.3em}\bigg)^2\bigg)\hspace{-0.3em}\bigg)}\Bigg]. \label{next_jit_exp}
\end{IEEEeqnarray}} \vspace{-0.5em}
\hrule \vspace{-0.5em} \end{figure*}
It is also evident from \eqref{next_jit_exp} that when there are no variations in interference and V2V distance, the associated jitter state evolution is zero. Further, jitter state evolution after a small period $(t - t^-)$ influenced by random deterioration processes can be mathematically derived as (from \eqref{state_dist_int}--\eqref{state_interfer}, and \eqref{next_jit_exp}) shown in \eqref{del_jit_dist} and \eqref{del_jit_int}.
\begin{figure*}[t!] \vspace{-1.5em}
\smaller{\begin{IEEEeqnarray}{rCl}	
\frac{\partial \tau_{\mathrm{tr}}(t)}{\partial \varsigma_d}
&=& \frac{p_I p_d L_p}{B}  \frac{\Bigg(\frac{P h^2 \lambda_{\textrm{op}}^2}{8 \pi^2 \varsigma_d^3(t) (N_0+ \zeta_I(t))}\Bigg)}{\ln(2) \bigg(1+\bigg(\frac{P h^2}{N_0+ \zeta_I(t)} \bigg(\frac{\lambda_{\textrm{op}}}{4\pi \varsigma_d(t)}\bigg)^2\bigg)\bigg) \Bigg(\log_2\bigg(1+\bigg(\frac{P h^2}{N_0+ \zeta_I(t)} \bigg(\frac{\lambda_{\textrm{op}}}{4\pi \varsigma_d(t)}\bigg)^2\bigg)\bigg)\Bigg)^2} + \frac{(1-p_I) p_d L_p}{B}  \nonumber\\&\times& \frac{\Bigg(\frac{P h^2 \lambda_{\textrm{op}}^2}{8 \pi^2 \varsigma_d^3(t) (N_0+ \overline{\zeta}_{I}(t^-))}\Bigg)}{\ln(2) \bigg(1+\bigg(\frac{P h^2}{N_0+ \overline{\zeta}_{I}(t^-)} \bigg(\frac{\lambda_{\textrm{op}}}{4\pi \varsigma_d(t)}\bigg)^2\bigg)\bigg) \Bigg(\log_2\bigg(1+\bigg(\frac{P h^2}{N_0+ \overline{\zeta}_{I}(t^-)} \bigg(\frac{\lambda_{\textrm{op}}}{4\pi \varsigma_d(t)}\bigg)^2\bigg)\bigg)\Bigg)^2}=p_I p_d f_1(P,N)+(1-p_I)p_d f_2(P,N). \label{del_jit_dist}\vspace{-0.3em}
\end{IEEEeqnarray}} \vspace{-0.5em}
\hrule \vspace{-0.5em}
\smaller{\begin{IEEEeqnarray}{rCl}	
\frac{\partial \tau_{\mathrm{tr}}(t)}{\partial \zeta_I}
&=& \frac{p_I p_d L_p}{B}  \frac{\Bigg(P\big(\frac{h\lambda_{\textrm{op}}}{4\pi \varsigma_d(t) (N_0+ \zeta_I(t))}\big)^2-1\Bigg)}{\ln(2) \bigg(1+\bigg(\frac{P h^2}{N_0+ \zeta_I(t)} \bigg(\frac{\lambda_{\textrm{op}}}{4\pi \varsigma_d(t)}\bigg)^2\bigg)\bigg) \Bigg(\log_2\bigg(1+\bigg(\frac{P h^2}{N_0+ \zeta_I(t)} \bigg(\frac{\lambda_{\textrm{op}}}{4\pi \varsigma_d(t)}\bigg)^2\bigg)\bigg)\Bigg)^2} + \frac{p_I (1-p_d) L_p}{B}  \nonumber\\&\times&\frac{\Bigg(P\big(\frac{h\lambda_{\textrm{op}}}{4\pi \overline{\varsigma}_d(t^-)(N_0+ \zeta_I(t))}\big)^2-1\Bigg)}{\ln(2) \bigg(1+\bigg(\frac{P h^2}{N_0+ \zeta_I(t)} \bigg(\frac{\lambda_{\textrm{op}}}{4\pi \overline{\varsigma}_d(t^-)}\bigg)^2\bigg)\bigg) \Bigg(\log_2\bigg(1+\bigg(\frac{P h^2}{N_0+ \zeta_I(t)} \bigg(\frac{\lambda_{\textrm{op}}}{4\pi \overline{\varsigma}_d(t^-)}\bigg)^2\bigg)\bigg)\Bigg)^2}=p_I p_d f_{1'}(P,N)+(1-p_d)p_I f_{2'}(P,N).\label{del_jit_int}
\end{IEEEeqnarray}} \vspace{-0.5em}
\hrule \vspace*{-2.3em}\end{figure*} 
Now we consider the scenario when both the impact of interference and V2V distance variations are present while the jitter state is evolving. Thereby, the mean and variance of the jitter state change after each transition can be derived approximately following the first-order Delta method using Taylor expansion \cite{delta} as follows:\vspace{-0.5em}
\begin{IEEEeqnarray}{rCl}	
\hspace{-1em}&\mathbb{E}&[\tau_{\mathrm{tr}}(t^+)] \approx \mathbb{E}[\tau_{\mathrm{tr}}(t)] \hspace{-0.2em}+ \hspace{-0.2em}\varepsilon_I \varepsilon_d\bigg(\frac{\partial \tau_{\mathrm{tr}}(t)}{\partial \zeta_I} \mu_{\zeta_I} \hspace{-0.3em}+\hspace{-0.2em} \frac{\partial \tau_{\mathrm{tr}}(t)}{\partial \varsigma_d} \mu_{\varsigma_d}\hspace{-0.3em}\bigg)\nonumber\\&+&\varepsilon_d(1-\varepsilon_I)\frac{\partial \tau_{\mathrm{tr}}(t)}{\partial \varsigma_d} \mu_{\varsigma_d} + \varepsilon_I(1-\varepsilon_d)\frac{\partial \tau_{\mathrm{tr}}(t)}{\partial \zeta_I} \mu_{\zeta_I},
\label{mean_del_jit_next}\vspace{-0.4em}
\end{IEEEeqnarray}
 where $\varepsilon_I$ and $\varepsilon_d$ are Bernoulli distributed random variables that capture the presence or absence of the interference and V2V distance variations, respectively, in the considered V2V link, such as $\Pr(\varepsilon_I=1)=p_I$, $\Pr(\varepsilon_I=0)=1-p_I$, $\Pr(\varepsilon_d=1)=p_d$, and $\Pr(\varepsilon_d=0)=1-p_d$. $\tau_{\mathrm{tr}}(t)$ is the transmission delay jitter at the previous time instant before the next evolved jitter state at $t^+$. Thus, it acts as the reference baseline for the newly evolved jitter state after each transition. Also, the variance of the jitter state change is written as:\vspace{-0.7em}
\begin{IEEEeqnarray}{rCl}	
&\mathbb{V}&[\tau_{\mathrm{tr}}(t^+)] \approx
\varepsilon_I \varepsilon_d\Bigg[\bigg(\frac{\partial \tau_{\mathrm{tr}}(t)}{\partial \zeta_I}\bigg)^2 \sigma_{\zeta_I}^2 + \bigg(\frac{\partial \tau_{\mathrm{tr}}(t)}{\partial \varsigma_d}\bigg)^2 \sigma_{\varsigma_d}^2\Bigg] \nonumber\\&\hspace{-1em}+& \hspace{-0.3em}\varepsilon_d (1\hspace{-0.2em}-\hspace{-0.2em}\varepsilon_I\hspace{-0.2em}) \bigg(\hspace{-0.3em}\frac{\partial \tau_{\mathrm{tr}}(t)}{\partial \varsigma_d}\hspace{-0.3em}\bigg)^2 \hspace{-0.3em}\sigma_{\varsigma_d}^2 \hspace{-0.2em}+\hspace{-0.2em} \varepsilon_I (1\hspace{-0.2em}-\hspace{-0.2em}\varepsilon_d) \bigg(\frac{\partial \tau_{\mathrm{tr}}(t)}{\partial \zeta_I}\bigg)^2 \sigma_{\zeta_I}^2.
\label{var_del_jit_next}\vspace{-0.5em}
\end{IEEEeqnarray}
It can be observed from \eqref{mean_del_jit_next} and \eqref{var_del_jit_next} that the mean and variance of each jitter state evolution are functions of the random, time-varying intensities of both deterioration processes. Also, each subsequent jitter state depends on the transmission delay profile of the previous packet transmission and the previous state of deterioration. As a result, the jitter state evolution follows a non-homogeneous semi-Markov process. Therefore, the generalized expression of the jitter intolerance probability of the considered V2V link at a particular time instant (shown in \eqref{jit_intol}) is analytically derived from \eqref{jit_intol} and expressed in \eqref{jit_intol_erf}, where $\text{erf}(\cdot)$ is the error function.
\begin{figure*}[t!]\centering
\vspace{-0.2em}\smaller{\begin{IEEEeqnarray}{rCl}	
\mathcal{P}_{JI}^+(t) 
&=&\underbrace{\frac{p_I p_d}{2} \Bigg[1+\text{erf}\underbrace{\Bigg(\frac{\frac{\partial \tau_{\mathrm{tr}}(t)}{\partial \zeta_I} \mu_{\zeta_I} \hspace{-0.3em}+\hspace{-0.2em} \frac{\partial \tau_{\mathrm{tr}}(t)}{\partial \varsigma_d} \mu_{\varsigma_d}\hspace{-0.3em}}{\sqrt{\bigg(\frac{\partial \tau_{\mathrm{tr}}(t)}{\partial \zeta_I}\bigg)^2 2 \sigma_{\zeta_I}^2 + \bigg(\frac{\partial \tau_{\mathrm{tr}}(t)}{\partial \varsigma_d}\bigg)^2 2 \sigma_{\varsigma_d}^2}}\Bigg)}_{x_1}\Bigg]}_{y_1} + \underbrace{\frac{(1-p_I)p_d}{2} \Bigg[1+\text{erf}\underbrace{\Bigg(\frac{ \frac{\partial \tau_{\mathrm{tr}}(t)}{\partial \varsigma_d} \mu_{\varsigma_d}\hspace{-0.3em}}{\sqrt{\bigg(\frac{\partial \tau_{\mathrm{tr}}(t)}{\partial \varsigma_d}\bigg)^2 2 \sigma_{\varsigma_d}^2}}\Bigg)}_{x_2}\Bigg]}_{y_2} \nonumber\\&+& \underbrace{\frac{(1-p_d)p_I}{2} \Bigg[1+\text{erf}\underbrace{\Bigg(\frac{\frac{\partial \tau_{\mathrm{tr}}(t)}{\partial \zeta_I} \mu_{\zeta_I}\hspace{-0.3em}}{\sqrt{\bigg(\frac{\partial \tau_{\mathrm{tr}}(t)}{\partial \zeta_I}\bigg)^2 2 \sigma_{\zeta_I}^2}}\Bigg)}_{x_3}\Bigg]}_{y_3}.\label{jit_intol_erf}
\end{IEEEeqnarray}} \vspace{-0.5em} \hrule \vspace{-2.em}
\end{figure*} 
Next, we derive the PDF of $\mathcal{P}_{JI}^+(t)$. To this end, we first obtain the PDF of $y_1$ utilizing the change of variables method with the transformation $x_1 =\mathrm{erf}^{-1}\left(\frac{2y_1}{p_I p_d} - 1\right)$ \cite{papoulis02}. Since $\mathrm{erf}(x_1) \in (-1,1)$, it follows that $y_1 \in (0, p_I p_d)$. Finally, $f_{Y_1}(y_1)$ can be computed as:

\vspace{-1.4em}
{\smaller\begin{IEEEeqnarray}{rCl}
 \hspace{-0.3em}f_{Y_1}(y_1) &\hspace{-0.3em}=\hspace{-0.3em}& f_{X_1}\hspace{-0.3em}\bigg(\hspace{-0.3em}\text{erf}^{-1}\hspace{-0.3em}\bigg(\hspace{-0.1em}\frac{2y_1}{p_I p_d}\hspace{-0.3em}-\hspace{-0.3em}1\hspace{-0.3em}\bigg)\hspace{-0.3em}\bigg) \frac{\sqrt{\pi}}{p_I p_d} \text{exp}\bigg(\hspace{-0.3em}\bigg(\hspace{-0.3em}\text{erf}^{-1}\hspace{-0.3em}\bigg(\hspace{-0.1em}\frac{2y_1}{p_I p_d}-1\bigg)\hspace{-0.3em}\bigg)^2\bigg). \label{Y1}
\end{IEEEeqnarray} }

\noindent Similarly, $f_{Y_2}(y_2)$ and $f_{Y_3}(y_3)$ can also be computed as follows:
\vspace{-1.5em}
{\smaller\begin{IEEEeqnarray}{rCl}
f_{Y_2}(y_2) &=& f_{X_2}\bigg(\text{erf}^{-1}\bigg(\frac{2y_2}{(1-p_I) p_d}-1\bigg)\bigg) \frac{\sqrt{\pi}}{(1-p_I) p_d} \nonumber\\&\times& \text{exp}\bigg(\bigg(\text{erf}^{-1}\bigg(\frac{2y_2}{(1-p_I) p_d}-1\bigg)\bigg)^2\bigg), \label{Y2}\vspace{-0.7em}
\end{IEEEeqnarray} }

\noindent and \vspace{-0.7em}
{\smaller\begin{IEEEeqnarray}{rCl}
f_{Y_3}(y_3) &=& f_{X_3}\bigg(\text{erf}^{-1}\bigg(\frac{2y_3}{p_I (1-p_d)}-1\bigg)\bigg) \frac{\sqrt{\pi}}{p_I (1-p_d)} \nonumber\\&\times& \text{exp}\bigg(\bigg(\text{erf}^{-1}\bigg(\frac{2y_3}{p_I (1-p_d)}-1\bigg)\bigg)^2\bigg). \label{Y3} \vspace{-0.6em}
\end{IEEEeqnarray} }

\noindent Therefore, the final PDF of jitter intolerance probability, including all the possible events, can be deduced as the triple convolution of PDFs $(\mathcal{P}_{JI}^+(t)=y_1+y_2+y_3)$ as: 

\vspace{-1em}
{\smaller\begin{IEEEeqnarray}{rCl}
&f&_{\mathcal{P}_{JI}^+(t)}(y) = (f_{Y_1}*f_{Y_2}*f_{Y_3})(y) \nonumber\\&=& \int\int f_{Y_1}(w) f_{Y_2}(z) f_{\mathcal{P}_{JI}^+(t)}(y-w-z)\text{d}w\text{d}z.\label{Y} \vspace{-0.7em}
\end{IEEEeqnarray} }

\noindent Similar to the random variable $y_1$, the supports of the other random variables are $y_2\in (0,(1-p_I) p_d)$ and $y_3\in (0,p_I (1-p_d))$. Thus, the lower and upper limits of $y$ can be computed as 0 and $p_I+p_d-p_I p_d$ $(=p_I p_d + (1-p_I) p_d + (1-p_d) p_I)$, respectively. As observed from \eqref{Y1}, \eqref{Y2}, and \eqref{Y3}, the individual PDFs cannot be convolved in a closed-form expression. Therefore, Laplace approximations are applied to capture the local behavior (near the mode) of each density. The convolution is then performed on the resulting approximate expressions. 
The detailed derivations are provided in Appendix A corresponding to the Laplace approximations. Finally, implementing the convention for the convolution of three independent Gaussian random variables, the approximate closed-form expression for the resulting PDF as a function of time is given by: \vspace{-0.5em}
\begin{IEEEeqnarray}{rCl} 
f_{Y}(y;t) =\frac{1}{\sqrt{2\pi\sigma_{y}(t)^2}}e^{\frac{-(y-y_{0}(t))^2}{2\sigma_{y}(t)^2}}, \vspace{-0em}\label{Y_closed} \vspace{-0.5em}
\end{IEEEeqnarray}
where $\sigma_{y}(t)^2=\sigma_{y_1}(t)^2+\sigma_{y_2}(t)^2+\sigma_{y_3}(t)^2$ and $y_{0}(t)=y_{10}(t)+y_{20}(t)+y_{30}(t)$.
In this section, $\mathbb{E}[\tau_{\mathrm{tr}}(t^+)]$ and $\mathbb{V}[\tau_{\mathrm{tr}}(t^+)]$ are shown to depend on $\varsigma_d(t)$ and $\zeta_I(t)$. Specifically, the growth of the V2V system's jitter statistics is governed by both the incremental evolution of the stressor distributions and the sensitivity of the jitter due to the considered stressors, $\frac{\delta \tau_{\mathrm{tr}}(t)}{\delta \varsigma_d}$ and $\frac{\delta \tau_{\mathrm{tr}}(t)}{\delta \zeta_I}$. This arises from the time-dependent nature of both the stressor states and $\mathcal{P}_{JI}^+(t)$, which captures the previous jitter condition at $t$ as a baseline and quantifies the likelihood of degradation in the subsequent state $t^+$. The resulting consolidated PDF $f_{\mathcal{P}_{JI}^+(t)}(y)$ characterizes the dynamic behavior of jitter intolerance, where the rate of jitter state transition depends on both the current stressor realization and the previous system state, thereby justifying a state-dependent Markov modeling framework. Moreover, the evolution remains influenced by controllable system parameters such as transmit power, operating wavelength, and the number of available wireless channels.

\vspace{-0.6em}
\section{Risk-based Jitter Performance Analysis}\vspace{-0.4em}
This section characterizes the temporal evolution of $\mathcal{P}_{JI}^+(t)$ derived in Section III, thus establishing its relationship with V2V communication performance, stressor-induced risk, and the safety operating boundaries of the corresponding V2V link. Accordingly, we introduce performance metrics to quantify both the evolution of jitter intolerance and the associated risk in the considered V2V system. As deterioration mechanisms intensify, the system's jitter performance falls below predefined safety thresholds, thereby increasing the susceptibility of the V2V link to transmission delay jitter. Given the inherently stochastic nature of the system, and to systematically incorporate resilience considerations into the jitter performance analysis, we develop probabilistic measures that characterize the V2V system's jitter behavior across the distinct phases of the resilience cycle, namely the alarming, failure, and restoration phases. The evolution rate of the jitter state for the considered V2V link driven by interference and inter-vehicular distance variation is quantified from \eqref{next_jit} and \eqref{jit_intol} by a novel resilience metric, the \textit{average risk exposure rate} (ARER). This metric characterizes the degradation rate (i.e., the slope of each jitter-state transition) of the jitter intolerance probability with respect to time, capturing how rapidly the probability deteriorates across successive time increments and along the evolving jitter states: 

\vspace{-1.5em}
{\smaller \begin{IEEEeqnarray}{rCl} 
\mathcal{E}_r(t^+) = \frac{\mathcal{P}_{JI}^+(t) -\mathcal{P}_{JI}^-(t)}{2 \Delta}, \label{ARER} \vspace{-0.5em}
\end{IEEEeqnarray} } 

\noindent where $(\cdot)^-$ and $(\cdot)^+$ correspond to the jitter intolerance probabilities during the jitter state transitions from $t^-$ to $t$ (during this transition the jitter state evolution is denoted as $\tau_{\mathrm{tr}}^-=d_{\mathrm{tr}}(t)-d_{\mathrm{tr}}(t^-)$), and from $t$  to $t^+$ (also, $ \tau_{\mathrm{tr}}^+=d_{\mathrm{tr}}(t^+)-d_{\mathrm{tr}}(t)$), respectively. $\Delta$ represents a single transition period ($t^-$ to $t$ or $t$ to $t^+$). In the presence of a continuously increasing stressor intensity, it is now essential to evaluate the system's jitter tolerance performance, which can quantify the ability of the V2V system to resist the increasing jitter risk. As a first step, we theoretically formulate the stepwise average jitter-withstanding capacity and the corresponding jitter-induced load across consecutive jitter state transitions of the considered V2V system as:

\vspace{-1.7em}
{\smaller \begin{IEEEeqnarray}{rCl}
\underbrace{\mathcal{C}(t^+)}_{\text{Jitter-withstanding capacity}}
= 1-\frac{\mathcal{P}_{JI}^+(t) +\mathcal{P}_{JI}^-(t)}{2}, \label{C}
\\
\underbrace{\mathcal{D}(t^+)}_{\text{Jitter-induced load}}
= \frac{\mathcal{P}_{JI}^+(t) +\mathcal{P}_{JI}^-(t)}{2}. \label{D} \vspace{-1em}
\end{IEEEeqnarray}}
\begin{figure}[t]
\vspace{-0.2em}\centerline{\psfig{file=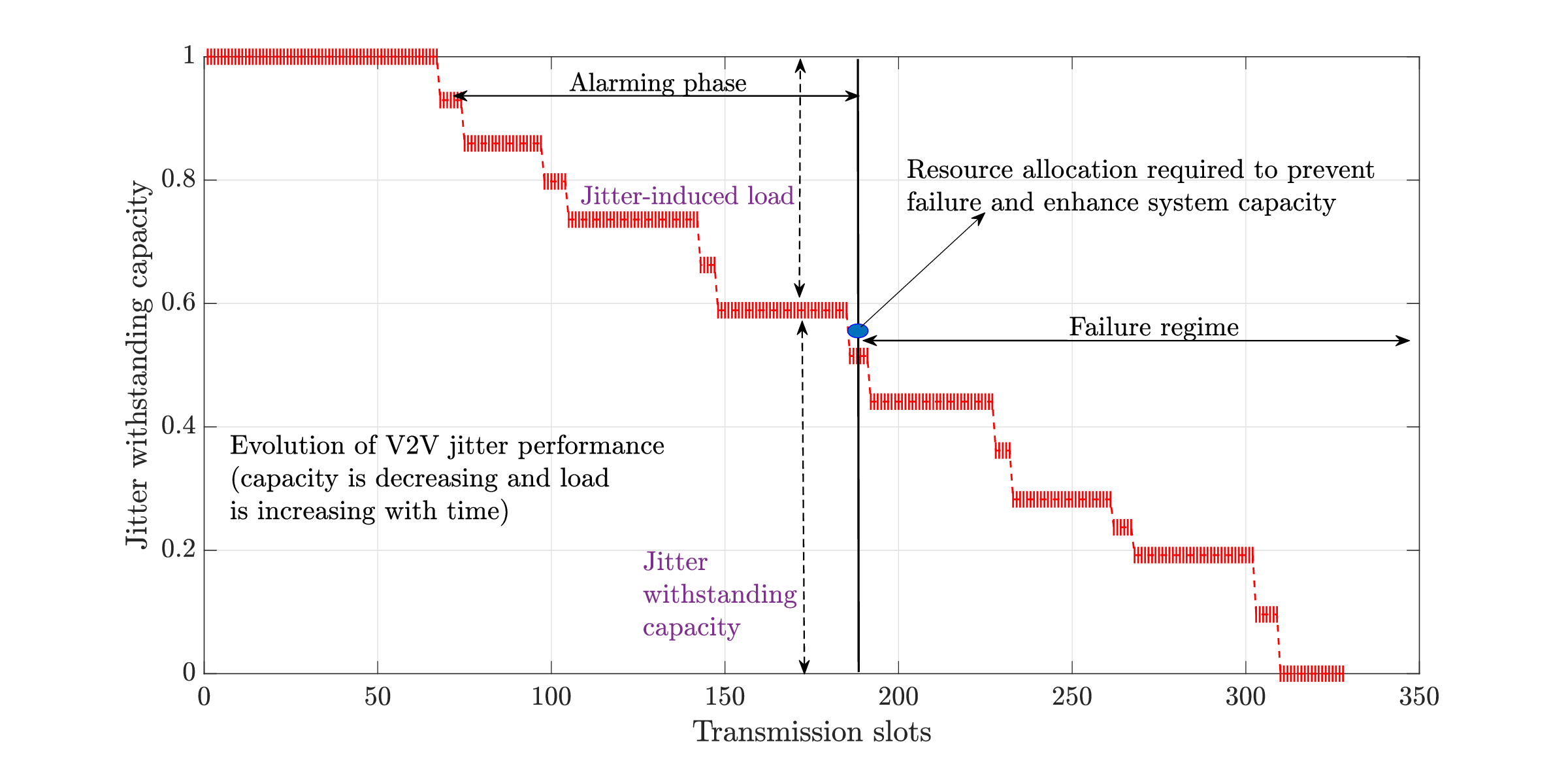,width=3.8in, height = 1.8 in }}
	\vspace{-0.7em} \caption{\small Impact of multiple stochastic deterioration processes on system performance.}
\vspace{-1.5em}
     \label{fig4}
\end{figure}

\noindent \eqref{C} and \eqref{D} represent the successive temporal evolution of the considered system's jitter-withstanding capacity and jitter-induced load as a function of $\mathcal{P}_{JI}$. As illustrated in Fig.~\ref{fig4}, we interpret the system's \textit{jitter-withstanding capacity} 
$\mathcal{C}(t^+)$ at any instant as the average jitter intolerance probability that the V2V link can still deliver while the jitter state evolves from the previous slot to the next in the presence of disturbances. During the alarming phase \textcolor{black}{(the regime where $\mathcal{P}_{JI}$ of V2V link increases but does not yet result in failure)}, $\mathcal{C}(t^+)$ decreases monotonically with
time. Conversely, the \textit{jitter-induced load} $\mathcal{D}(t^+)$ is the
\textit{average jitter intolerance performance loss} accumulated over that same slot transition; it grows monotonically throughout the alarming stage.  
To this end, \textit{a limit-state function \cite{paolo} is introduced as a V2V system's jitter state indicator, based on the relationship between the jitter-induced load and the V2V system's jitter-withstanding capacity}. This function enables the identification of threshold-crossing events where jitter performance begins to degrade. It can also be observed from Fig.~\ref{fig4} that in the absence of interference and V2V distance variations, both the capacity and load dynamics remain $\mathcal{C}(t^+)\approx 1$ and
$\mathcal{D}(t^+)\approx 0$.
As deterioration progresses, capacity falls toward~$0$ while load rises toward~$1$, so the limit-state function can collectively track both trends to closely incorporate the impacts of stochastic deterioration as well as the system's jitter-withstanding capacity. Next, we develop a resilience metric in the form of a limit-state indicator function that quantifies the system's ability to withstand jitter over time.  Therefore, the instantaneous limit-state can be defined as:

\vspace{-1.3em}
{\small \begin{IEEEeqnarray}{rCl}
\hspace{-1em}\mathcal{G}(t^+) = \mathcal{C}(t^+) - \mathcal{D}(t^+) =1-\big(\mathcal{P}_{JI}^+(t) +\mathcal{P}_{JI}^-(t)\big).\label{lim}\vspace{-.7em}
\end{IEEEeqnarray}}

\noindent Intuitively, the limit-state indicator $\mathcal{G}(t^+)$ quantifies the margin between the system's remaining capacity to withstand jitter and the accumulating jitter load. A positive value of $\mathcal{G}(t^+)$ indicates the system is still within safe operational limits, while a transition to negative signifies that jitter-induced load has exceeded the system's capacity, marking the onset of critical performance degradation. If jitter intolerance probability continues to grow, driven by mobility and interference-induced factors, it raises a critical concern about the system's ability to maintain operational stability. For this reason, in the next stage, we emphasize maintaining the stability margin of the associated limit-state function, which identifies the critical operating condition at which the system loses its ability to withstand the evolving jitter profile, and corrective action must be initiated for recovery.

\vspace{-.5em}
\section{Resource Allocation During Post-failure}
In this section, we focus on enhancing $\mathcal{C}(t)$ so that $\mathcal{G}(t)$ can return to its safe operational value. Let's consider the failure instant $t_f$, the limit-state $\mathcal{G}(t_f)$ falls to zero or below, which indicates that the link fails to maintain its jitter-withstanding capacity in the presence of the continuously evolving deterioration processes. At this stage, resilience is instilled into the system through the initiation of the restoration process by resource allocation. Although vehicles can communicate directly over V2V links, roadside unit (RSU) assistance is not required under normal operating conditions. In such a scenario, direct communication is generally sufficient to support timely message exchange with lower infrastructure involvement. However, when transmission-delay jitter degrades the communication conditions, the RSU can enhance communication resilience by supporting the affected V2V link through resource allocation \cite{rp, rp1} based on jitter variations. Moreover, building on RSU‑centric approaches \cite{sparse, rsu1, rsu2, rsu3}, including centralized orthogonal frequency division multiplexing scheduling \cite{optz1}, interference‑aware reuse schemes for V2V networks \cite{optz4}, and RSU‑assisted beamforming and spatial coordination \cite{optz2} in mmWave V2V systems, it is evident that RSUs can enforce time-frequency-space separation and controlled reuse among strongly interfering V2V links. Hence, RSUs suppress dominant co-channel interference and enable decoupled per-link power and rate optimization under reuse and interference constraints. From the aforementioned discussions, it is assumed in the considered V2V setup that the RSU is responsible for adaptive power allocation and the implementation of diversity mechanisms, while simultaneously controlling interference.

In this regard, using full power or a large number of transmit antennas from the initial phase may seem optimal for maximizing performance in the short term, but it is suboptimal and potentially harmful for long-term efficient and jitter-resilient V2V communication. Because in V2V systems operating under uncertain, time-varying, or adverse environmental conditions, such as interference and mobility-induced fading, using maximum resources can degrade the performance, in terms of jitter, of the wireless link. For instance, using higher transmit power without accounting for the jitter-induced load on the V2V link can increase the transmission delay jitter and make the link more vulnerable to jitter violations. In contrast, an adaptive resource allocation policy allows the system to utilize resources based on the current system state and uncertainty. This not only enhances resilience against severe jitter conditions but also helps maintain consistent system performance over time and supports adaptive jitter mitigation mechanisms. High resource utilization should be reserved for high-impact moments because resilience is not simply about always being strong, but about knowing when to be strong. Therefore, for the given system model, achieving a higher positive value of $\mathcal{G}(t_f^{+})$ during the restoration phase is desirable, as it minimizes the jitter intolerance probability in subsequent transmission slots. Two case studies are presented to demonstrate how different resource allocation policies can enhance $\mathcal{G}(t)$ under progressively deteriorating V2V conditions. In this context, to implement the MISO mechanism and enable adaptive power allocation, the instantaneous SINR for the considered V2V setup at each time slot $t$ is expressed as (from \eqref{SINR}):

\vspace{-1.5em}
{\smaller \begin{IEEEeqnarray}{rCl}
\gamma'(t)=\frac{P_q (\sum_{j=1}^{N_t}{h_j}^2/N_t)}{N_0+ \zeta_{I}(t)} \bigg(\frac{\lambda_{\textrm{op}}}{4\pi \varsigma_d(t)}\bigg)^2, \label{SINR_res} \vspace{-0.3em}
\end{IEEEeqnarray}}

\noindent where $\{h_1,h_2,..,h_j\}$ is an array of fading channel gains for the corresponding $N_t$ number of arbitrary transmission links at transmit vehicle end. The rest of the jitter formulation and also the deterioration behavior will follow similar mathematical steps as provided in Section II. We study the impact of each resource allocation scheme on the probability of jitter intolerance separately. 
\vspace{-4mm}
\subsection{Adaptive power allocation and diversity schemes}
Consider the restoration process initiates at $t=t_f$ and the failure instant is reset as $t_f=0$ when $\mathcal{G}\leq0$. The limit-state function $\mathcal{G}(t)$ is evaluated across $q,q'$ threshold levels ($q,q' \in \{1, 2, \ldots, m_{q,q'}\}$), where each level $\mathcal{G}_{q,q'}$ is associated with a corresponding time segment of $k_{q,q'}$ steps, $k_{q,q'} \in \{1, 2, \ldots, n_{k,k'}\}$, required for performance improvement. Each superscript $\{'\}$ represents the processes associated with the adaptive MISO scheme \footnote{Please note that multiple-input multiple-output (MIMO) V2V links are not considered in this work due to the additional modeling complexities they introduce under high mobility. In particular, MIMO operation requires receiver-side spatial combining and processing, which leads to directional and subspace-dependent interference coupling, increases synchronization and beam alignment sensitivity, and introduces spatial correlation effects at the destination vehicle, all of which substantially complicate interference management and system-level performance analysis.} with arbitrary $N_t$ V2V links. For resource allocation, the optimizable function during the restoration stage can be mathematically computed as a new metric, the probability of limit state exceedance, given as:

\vspace{-1em}
\textcolor{black}{{\smaller \begin{IEEEeqnarray}{rCl}
\hspace{-1.2em}\Pr(\mathcal{G}(t^+ + k_{q,q'} \Delta t)> 0)&=& \Pr(\mathcal{P}_{JI}^+(t)< (1-\mathcal{P}_{JI}^-(t))), \nonumber\\&=& \hspace{-0.3em}1\hspace{-0.3em}-\hspace{-0.5em}\int_{1-\mathcal{P}_{JI}^-(t)}^{p_I+p_d-p_I p_d} \hspace{-0.5em}f_{\mathcal{P}_{JI}^+(t)}\hspace{-0.1em}(y)\text{d}y.
\label{exd}\vspace{-1em}
\end{IEEEeqnarray}}}

\noindent The problem of allocating optimized power $P_q^*$ or MISO link $N_{t_{q'}}^*$ to achieve a certain performance level of $\mathcal{G}_{q,q'}$ in the presence of deterioration can be formulated as a $\{q,q'\}$-stage optimization problem. In this regard, a sequential recovery process is used in power and MISO link adaptation because wireless channels vary rapidly and unpredictably due to fading, interference, and mobility, making static strategies unreliable \cite{adapt1, adapt2}. This stepwise approach improves robustness and speeds recovery from link degradation compared to less frequent or bulk updates \cite{adapt3}. Frameworks such as LiBRA \cite{adapt4} show sequential processing balances throughput and recovery delay by triggering adaptations based on current conditions and history, while reducing estimation errors and adaptation instability. Thus, during each interval $[t^+,t^+ +k_{q,q'}\Delta t]$, the successive improvement of $\mathcal{G}_{q,q'}$ can be mathematically formulated in terms of an objective function following the expression of probability of limit state exceedance expressed in \eqref{exd}. To determine the optimal transmission power and transmit antenna scheduling configuration that maximizes the likelihood of successful packet transmission with minimal jitter, we formulate an optimization problem centered around minimizing the cumulative probability mass $(\Pr(\mathcal{P}_{JI}^+(t)> (1-\mathcal{P}_{JI}^-(t))))$ of a Gaussian random variable $ Y \sim \mathcal{N}(y_0, \sigma_y^2) $ within a dynamic interval $[a, b]$.
Here, the upper limit $b = p_I + p_d - p_I p_d$ is fixed, while the lower limit $a = 1 - \mathcal{P}_{JI}^-(t)$ depends on the dynamic deterioration-induced degradation probability $\mathcal{P}_{JI}^-(t)$, which in turn is influenced by the decision variables $\{P_q\}, \{N_{t_{q'}}\}$. The mean $y_0$ and variance $\sigma_y^2$ of the random variable $Y$ are also modeled as functions of these decision variables. Also, in the minimization of a certain event problem, the optimization is performed under uncertainty, where it is required that a certain constraint, such as $(\mathcal{P}_{JI}^+(t)< (1-\mathcal{P}_{JI}^-(t)))$, be satisfied with high probability rather than always. So this optimization problem can be formulated as a chance-constrained optimization problem. As in this problem,$(1-\mathcal{P}_{JI}^-(t))$ is the lower bound of the probabilistic support, so an acceptable threshold is introduced above that lower bound, which is $\theta(t)=(1-\mathcal{P}_{JI}^-(t))+\psi(t)$ with $0<\psi(t)<b-(1-\mathcal{P}_{JI}^-(t))$. Therefore, $(1-\mathcal{P}_{JI}^-(t))$ is the minimum feasible value, $\psi(t)$ is the margin above the minimum and $\theta(t)$ is the acceptable upper limit for $\mathcal{P}_{JI}^+(t)$. Since the need is to keep $\mathcal{P}_{JI}^+(t)$ to stay below that acceptable limit with high probability, it can be mathematically formulated in a chance constrained form as:

\vspace{-2.em}
{\smaller \begin{IEEEeqnarray}{rcl}
    \Pr(\mathcal{P}_{JI}^+(t) \leq 1 - \mathcal{P}_{JI}^-(t) + \psi(t)) \geq 1-\alpha, \label{cco}\vspace{-.5em}
\end{IEEEeqnarray}} 

\noindent where $\alpha$ is the maximum allowed violation probability or equivalently $1-\alpha$ is the minimum required confidence level. As the objective is to minimize $\alpha$ or to increase the confidence, the final chance constrained optimization problem can be formulated as:

\vspace{-1.4em}
\begin{subequations}
{\smaller \begin{align}
\hspace{-1em}\quad \min \,\,&\psi(t) 
\IEEEyesnumber\IEEEyessubnumber* \\[4pt]
\text{s.t.}\,& \Pr(\mathcal{P}_{JI}^+(t)+ \mathcal{P}_{JI}^-(t)-1 \leq \psi(t)) \geq 1-\alpha \label{eq:chance_const} 
            \\[2pt]
            & 0 \leq \mathcal{P}_{JI}^-(t) \leq 1 ,\label{eq:const_PJI1} 
\end{align} }
\label{eq_problem1}\vspace{-1.5em}
\end{subequations}

\noindent Since $\mathcal{P}_{JI}^{+}(t) \sim\mathcal{N}\bigl(y_0(t), \sigma_y^2(t)\bigr)$, \eqref{eq_problem1} can further be expanded as:

\vspace{-1.5em}
\begin{subequations}
{\smaller \begin{align}
\hspace{-1em}\quad \min \,\,&\psi(t) 
\IEEEyesnumber\IEEEyessubnumber* \\[4pt]
\text{s.t.}\,& \Phi\bigg(
\frac{\psi(t)+1-\mathcal{P}_{JI}^{-}(t)-y_0(t)}{\sigma_y(t)}
\bigg)\geq 1-\alpha  \label{eq:chance_const_form} 
\\[2pt] & 0 \leq P_q \leq P_q^{\max}, \quad \forall q \label{eq:const_power1} \\[2pt]
            &  \sum_q P_q \leq P_{\text{total}} \label{eq:const_budget1} , \\[2pt]
            & N_{t_{q'}} \in \mathbb{Z}_+, \quad \forall q' \label{eq:const_Nt1} 
            \\[2pt]
            & 0 \leq \mathcal{P}_{JI}^-(t) \leq 1 ,\label{eq:const_PJI1} 
            \\[2pt]
            & y_0(t) = f(\{P_q\}, \{N_{t_{q'}}\}), \\[2pt]
            & \sigma_y(t) = g(\{P_q\}, \{N_{t_{q'}}\}). 
 \label{eq:const_moments1}
\end{align}} 
\label{eq_problem2}\vspace{-1em}
\end{subequations}
\begin{figure}[t]
\vspace{-3mm}
\begin{algorithm}[H]
\caption{\small Proposed algorithm}
\label{algo}
\begin{algorithmic}[1]\small
\vspace{-2mm}
\State \textbf{Given:} Calculate $\mathcal{G}(t_f)$ and $\mathcal{E}_r(t_f)$ with the corresponding parameter value when $\mathcal{G}(t_f)<0$ and initialize the failing time instant $t=t_f$ when resource allocation $P_q$ or $N_{t_{q'}}$ starts\;
\For{each repetition step $\rightarrow$ each jitter state evolution} 
     \State Compute $\mathcal{G}(t^+)$
     \State \textbf{if} $\mathcal{G}(t^+)<1$, compute $\mathcal{E}_r(t)$ 
     \State \textbf{if} $\mathcal{E}_r(t^+)>\mathcal{E}_r(t^-)$, 
     $P_q$ or $N_{t_{q'}}$ increment in the presence of stressor uncertainty
     \State Compute $\mathcal{G}(t_f)$, $\mathcal{E}_r(t_f)$
   \State \textbf{else} Hold the current state
    \State \textbf{else} break
\EndFor
\end{algorithmic}
\label{Algo1}
\end{algorithm} 
\vspace{-10mm}
\end{figure}

\noindent Here, $f(\{P_q\}, \{N_{t_{q'}}\})$ and $g(\{P_q\}, \{N_{t_{q'}}\})$ 
denote the mean and standard deviation of $\mathcal{P}^+_{JI}(t)$ expressed in \eqref{Y_closed}, respectively, which are nonlinear functions of the decision variables $\{P_q\}$ and $\{N_{t_{q'}}\}$ through the jitter sensitivity 
expressions in~(31) and~(32) and the stressor statistics 
in~(23),~(24),~(26), and~(27). Fixing the integer variables $\{N_{t_{q'}}\}$, the chance constraint {\smaller $\Phi\bigg(
\frac{\psi(t)+1-\mathcal P_{JI}^{-}(t)-f(P_q)}{g(P_q)}
\bigg)\geq 1-\alpha$} is equivalent to
{\smaller$\psi(t)\ge f(P_q)+\Phi^{-1}(1-\alpha)g(P_q)+\mathcal P_{JI}^{-}(t)-1$}. Since $\psi(t)$ is minimized and appears only through the above lower bound, the inequality is active at the optimum. Hence, the reduced problem can be written as:

\vspace{-1.4em}
\begin{subequations}
{\smaller \begin{align}
\min_{P_q}\quad &
F(P_q):=f(P_q)+\Phi^{-1}(1-\alpha)g(P_q)
\\
\text{s.t.}\quad
& 0\le P_q\le P_q^{\max}, \quad \forall q,
\\
& \sum_q P_q\le P_{\mathrm{total}}. 
\end{align}}
\label{eq_power_allocation}
\end{subequations}\vspace{-.5em}

\noindent Accordingly, the Lagrangian can be defined as:
\vspace{-0.4em}
{\smaller\begin{IEEEeqnarray}{rcl}
 \mathcal{L}&=&F(P_q)
-\sum_q \Lambda_q P_q
+\sum_q \varrho_q(P_q-P_q^{\max})
\nonumber\\&+&\epsilon\big(\sum_q P_q-P_{\mathrm{total}}\big),\vspace{-0.5em}
\end{IEEEeqnarray}}
\hspace{-0.5em}where $\Lambda_q,\varrho_q,\epsilon\ge 0$. The stationary Karush-Kuhn-Tucker (KKT) conditions can now be formulated as:

\vspace{-1.3em}
{\smaller \begin{IEEEeqnarray}{rcl}
\hspace{-1em}\frac{\partial f(P_q)}{\partial P_q}
+\Phi^{-1}(1-\alpha)\frac{\partial g(P_q)}{\partial P_q}-\Lambda_q+\varrho_q+\epsilon=0,~ \forall q,\label{eq_KKT} \vspace{-0.7em}
\end{IEEEeqnarray}}

\noindent where these are solved together with {\smaller $0\le P_q\le P_q^{\max}, \forall q,~\sum_q P_q\le P_{\mathrm{total}}$, $\Lambda_q\ge 0,~ \varrho_q\ge 0,~ \epsilon\ge 0$,
and $\Lambda_q P_q=0,~
\varrho_q(P_q-P_q^{\max})=0,~
\epsilon\left(\sum_q P_q-P_{\mathrm{total}}\right)=0$}.
Therefore, the optimal solution satisfies:\vspace{-0.7em}
{\smaller {\begin{align}
 \frac{\partial f(P_q^\star)}{\partial P_q}
+\Phi^{-1}(1-\alpha)\frac{\partial g(P_q^\star)}{\partial P_q}
+\epsilon^\star
\begin{cases}
\ge 0, & P_q^\star=0,\\
=0, & 0<P_q^\star<P_q^{\max},\\
\le 0, & P_q^\star=P_q^{\max}.
\end{cases}\vspace{-0.7em}
\end{align}}}

\noindent Similarly, the optimization problem can also be solved for $N_{t_{q'}}$ while keeping $P_q$ fixed. \textcolor{black}{Please note that the transformation of \eqref{eq_problem2} to \eqref{eq_power_allocation} is exact and does not alter the underlying optimization problem. Hence, the solved problem is equivalent to the original formulation, and the derived solution remains optimal.} In addition to that, the optimal solution is determined over the feasible set induced by the deterministic equivalent of the chance constraint, in which uncertainty is characterized through the mean $y_0(t)$ and the standard deviation $\sigma_y(t)$. By reformulating the probabilistic constraint via the inverse Gaussian cumulative distribution function, the original chance-constrained problem is converted into a deterministic mean-risk optimization problem. Because $y_0(t)=f(\{P_q\},\{N_{t_{q'}}\})$ and $\sigma_y(t)=g(\{P_q\},\{N_{t_{q'}}\})$, the resulting formulation is generally nonlinear and may be nonconvex, depending on the specific forms of $f(\cdot)$ and $g(\cdot)$. Therefore, the KKT conditions provide the necessary first-order conditions for local optimality, provided that the relevant constraint conditions are satisfied. Thus, the optimized values of $P_q$ and $N_{t_{q}}$ can minimize larger jitter intolerance in the considered dynamic V2V system model. 
\begin{table}[t]
\centering
\caption{\textcolor{black}{Simulation parameters \cite{SNR_V2V, path_loss_V2V, interfer_vehicle}}}
 \begin{tabular}{|c|c|c|}
\hline
\textbf{Parameter} & \textbf{Description} & \textbf{Value} \\
\hline
$\mu$ & Deterministic drift of distance (m/s) & 0.2 \\
$v_0$ & Initial distance variance & $0.5^2$ \\
$P$ & Transmit power (dBm) & 30 \\
$P_j$ & Transmit power per interferer (dBm) & 5\\
$d_j$ & Random distance range of interferer (m) & [20 100] \\
$\lambda_{\textrm{op}}$ & Operating wavelength (for 5.9GHz) (m) & 0.05 \\
$L_p$ & Packets per transmission slot (bits) & 2400  \\
$B$ & Bandwidth (MHz) & 20 \\
$\alpha$ & Path loss exponent & 3.5 \\
$\lambda_0$ & Arrival rate (vehicles/s) & 0.2 \\
$N_0$ & Noise power spectral density (dBm/Hz) & -150\\
$I_0$ & Initial interferer count & 5 \\
$d_0$ & Initial distance (m) & 10 \\
$\eta$ & Growth‑rate of distance variance & 0.4 \\
$p_H$ & Sojourn time rate & $0.5$ \\
$\xi$ & Volatility of distance variance & 0.35 \\
$\omega$ & Linear interference growth factor & 0.8 \\
$\varphi$ & Non-linear interference growth factor & 0.7\\
\hline
\end{tabular}
\vspace{-2em}
\end{table} 
\begin{figure*}[t]
\vspace{-0.5em}
\centering

\begin{minipage}{0.32\textwidth}
    \centering
    \psfig{file=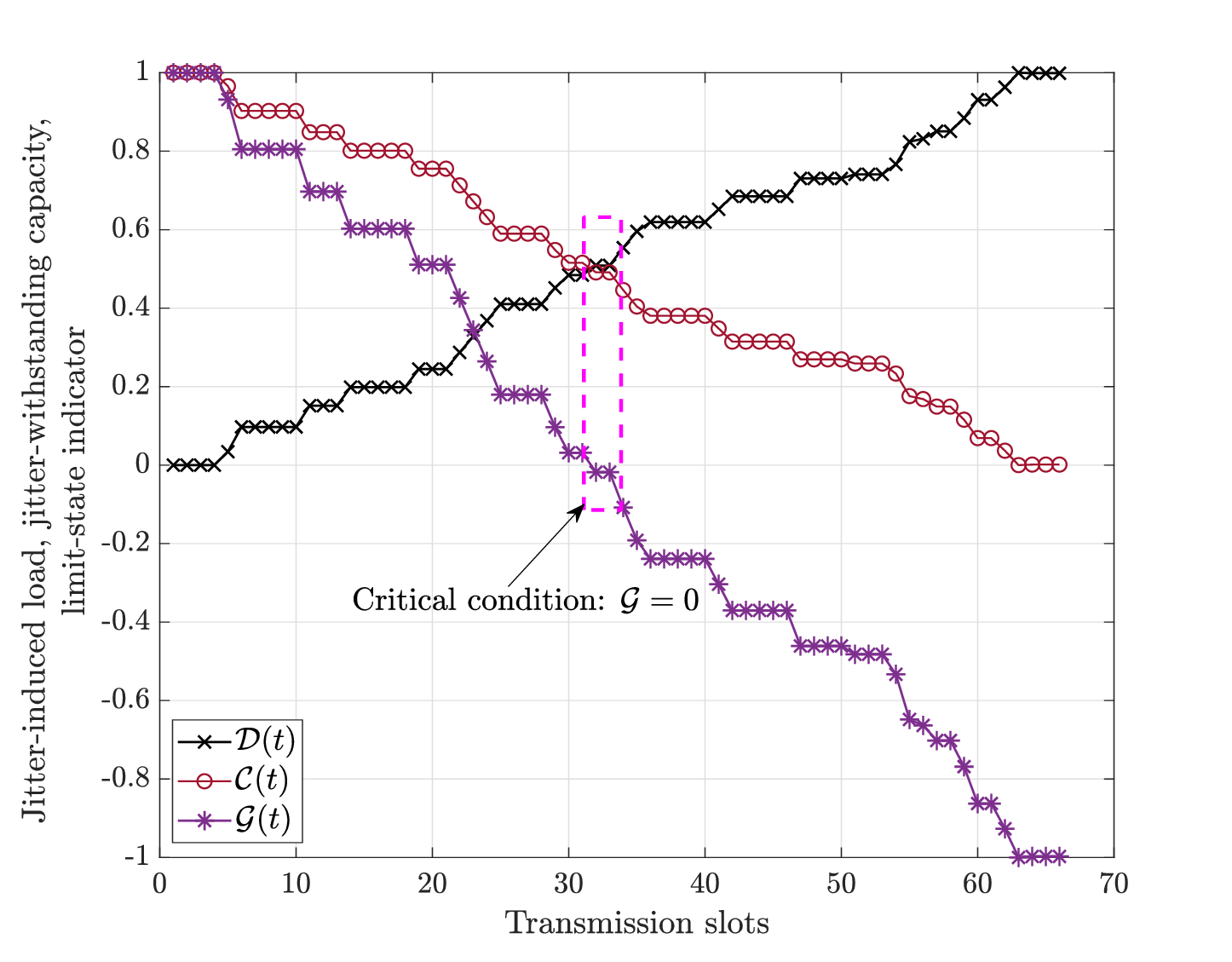,width=2.5in,height=1.9in}
    \vspace{-1em}
    
    {\vspace{-0.7em}\small Fig. 3: $\mathcal{C}(t)$, $\mathcal{D}(t)$, and $\mathcal{G}(t)$ due to only $\zeta_I$.} 
\end{minipage}
\hfill
\begin{minipage}{0.32\textwidth}
    \centering
    \psfig{file=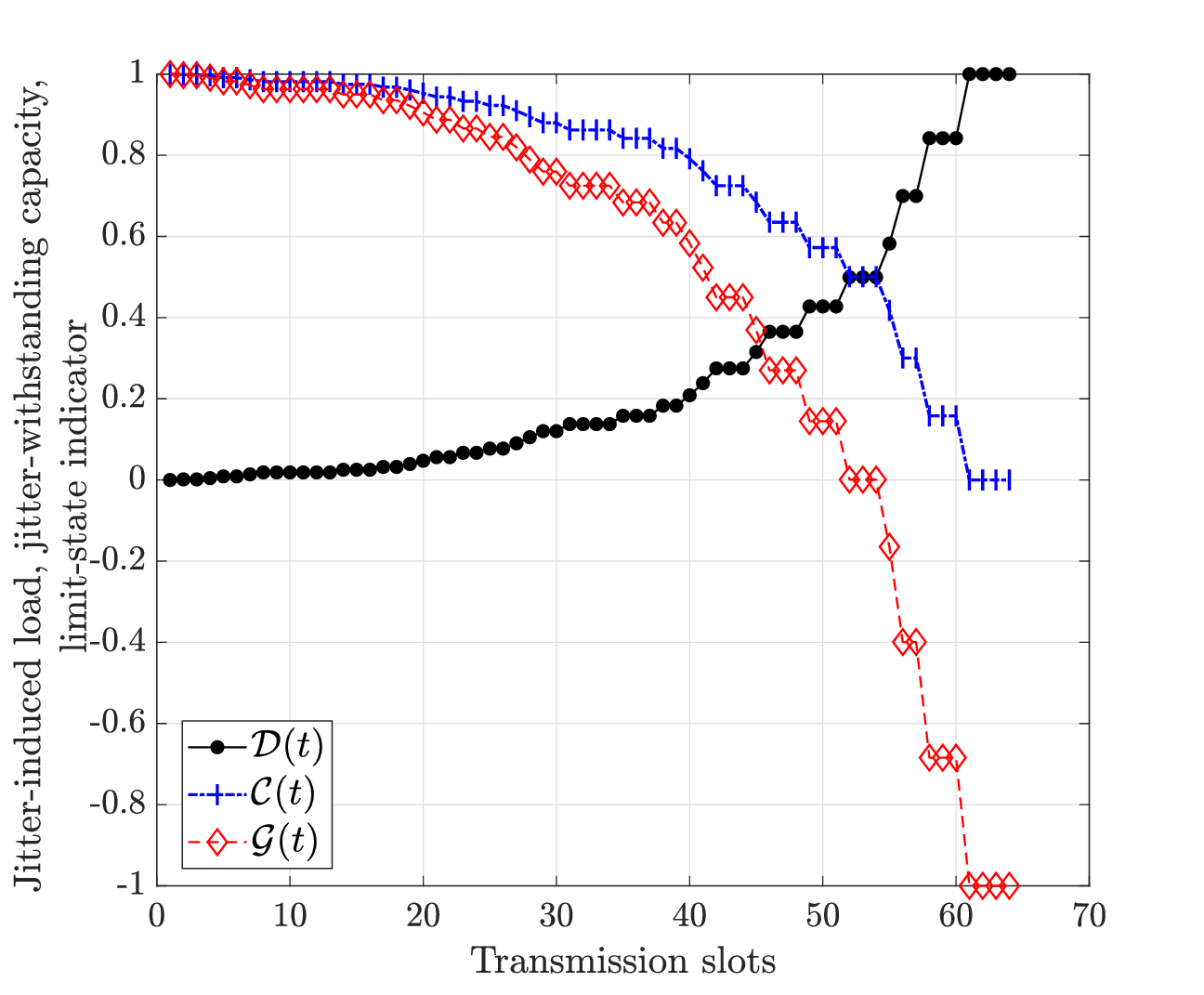,width=2.5in,height=1.9in}
    \vspace{-1em}
    
    {\vspace{-0.7em}\small Fig. 4: $\mathcal{C}(t)$, $\mathcal{D}(t)$, and $\mathcal{G}(t)$ due to only $\varsigma_d$.}
\end{minipage}
\hfill
\begin{minipage}{0.32\textwidth}
    \centering
    \psfig{file=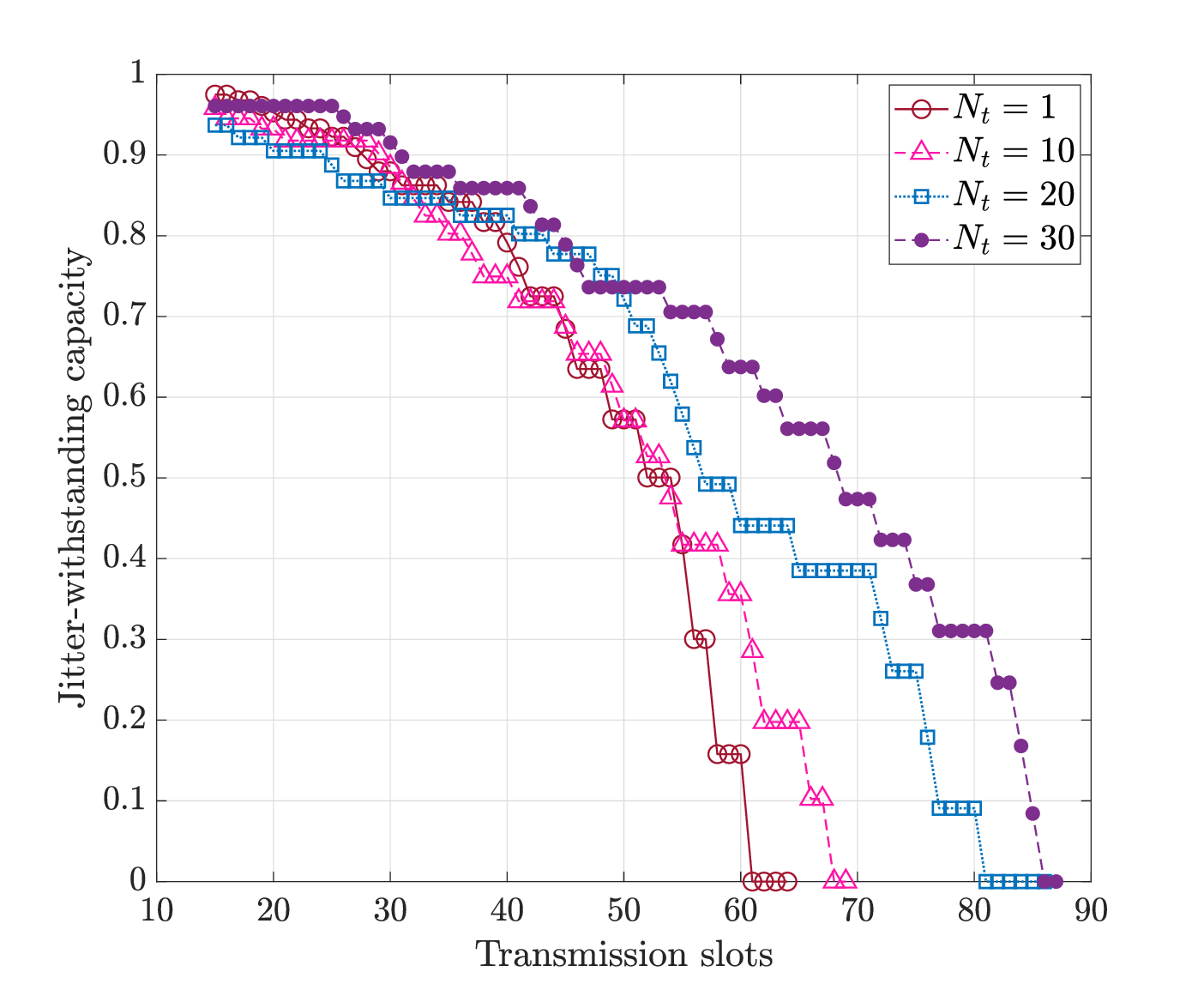,width=2.5in,height=1.9in}
    \vspace{-1em}
    
    {\vspace{-0.7em}\small Fig. 5: $\mathcal{C}(t)$ due to $\varsigma_d$ with arbitrary $N_t$.}
\end{minipage}
\vspace{-1.5em}
\end{figure*}
\begin{figure*}[t]
\vspace{.1em}
\centering

\begin{minipage}{0.32\textwidth}
    \centering
    \psfig{file=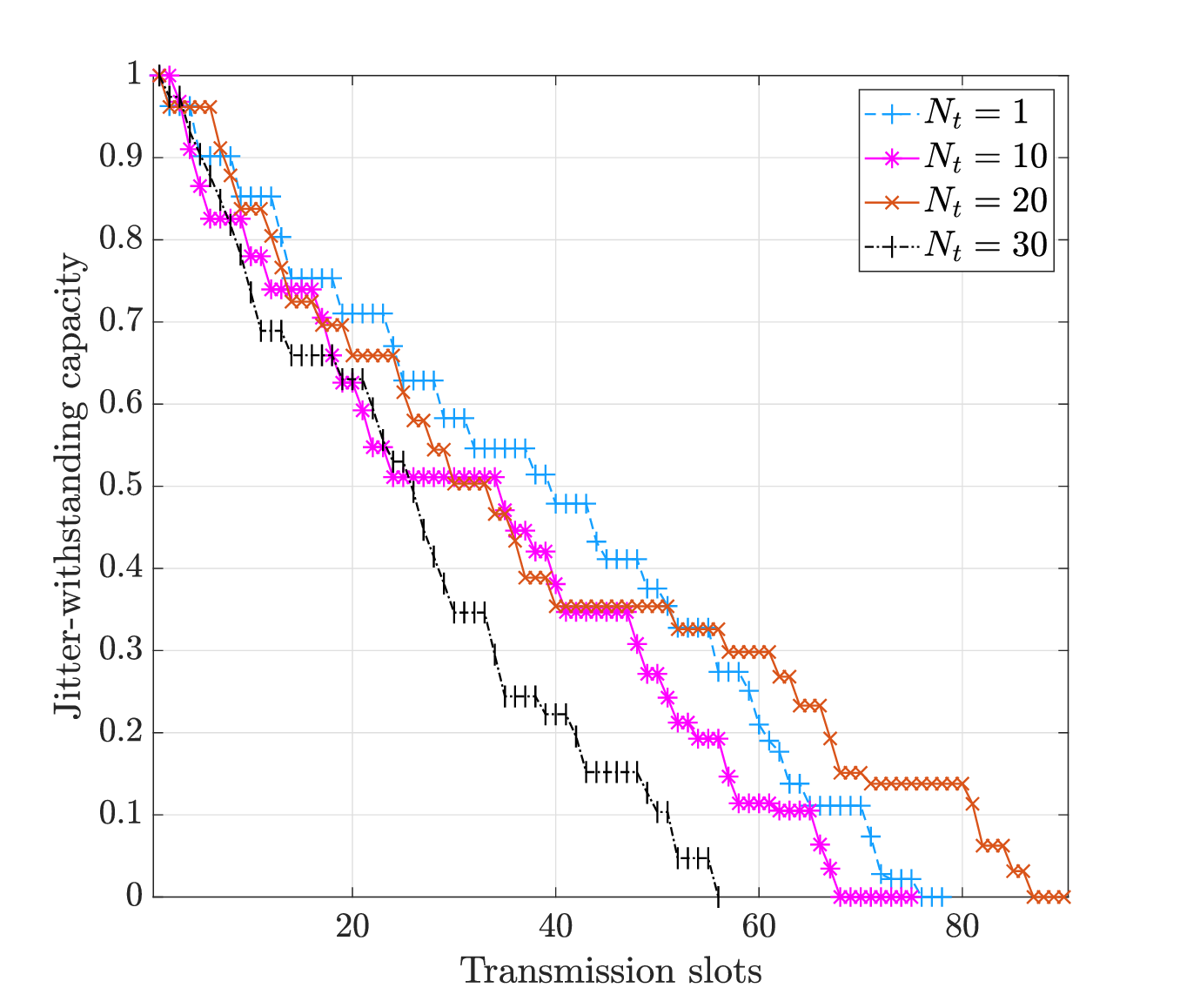,width=2.5in,height=1.9in}
    \vspace{-1em}
    
    {\vspace{-0.7em}\small Fig. 6: $\mathcal{C}(t)$ due to $\zeta_I$ with arbitrary $N_t$.}
\end{minipage}
\hfill
\begin{minipage}{0.32\textwidth}
    \centering
    \psfig{file=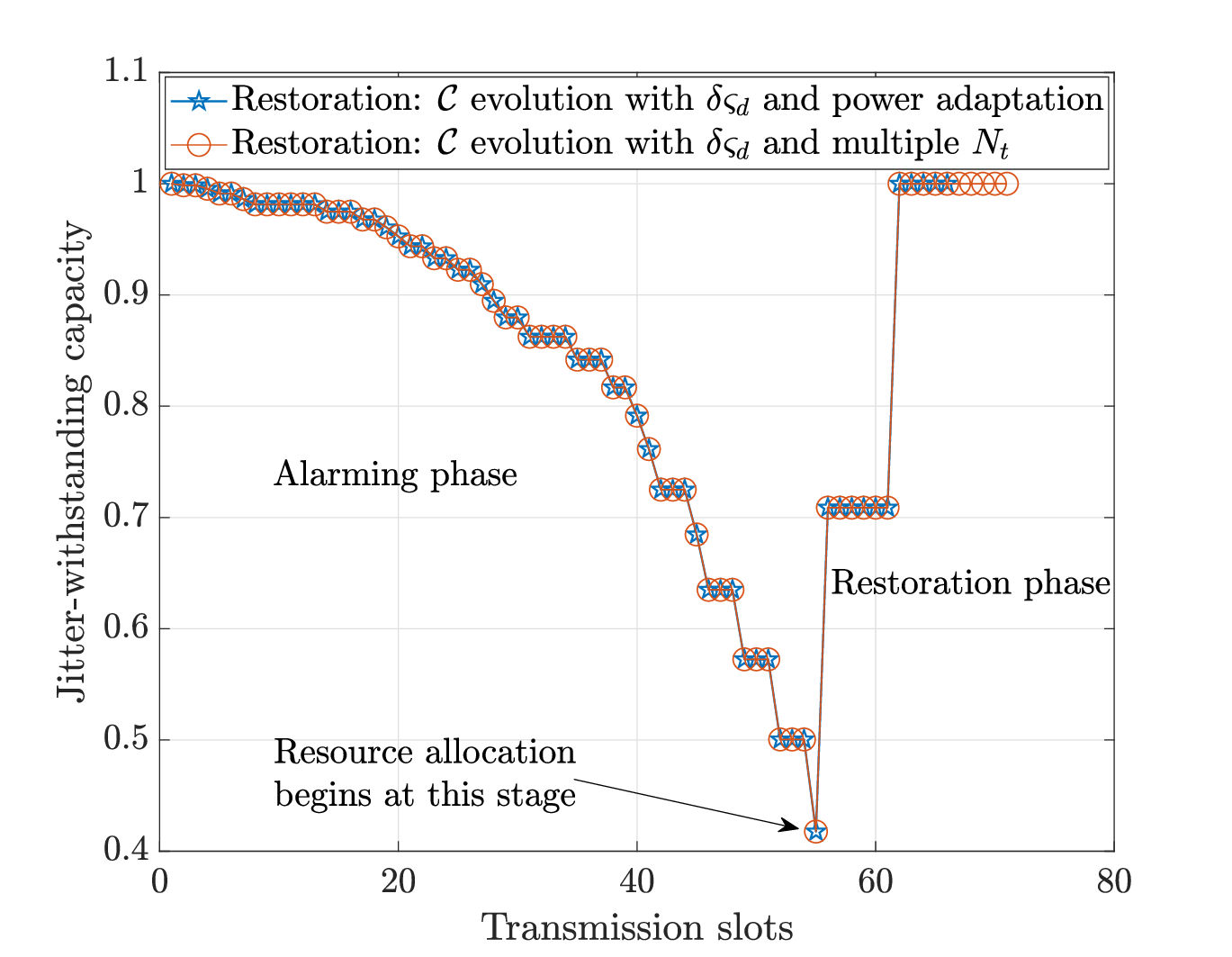,width=2.5in,height=1.9in}
    \vspace{-1em}
    
    {\vspace{-0.7em}\small Fig. 7: $\mathcal{C}(t)$ restoration in the presence of $\varsigma_d$.}
\end{minipage}
\hfill
\begin{minipage}{0.32\textwidth}
    \centering
    \psfig{file=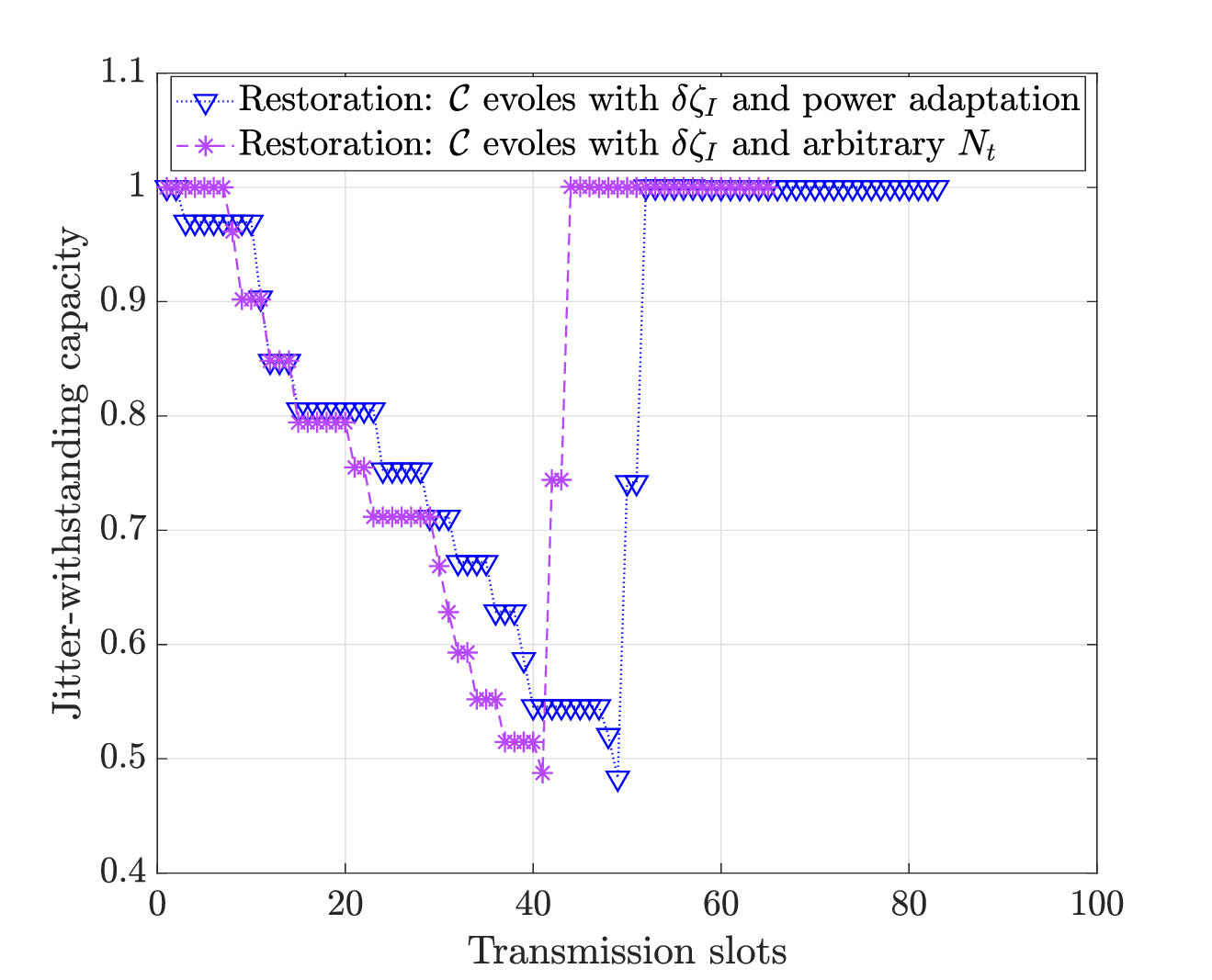,width=2.5in,height=1.9in}
    \vspace{-1em}
    
    {\vspace{-0.7em}\small Fig. 8: \hspace{-0.1em}$\mathcal{C}(t)$ restoration in the presence of $\zeta_I$.}

\end{minipage}
 \vspace{-1.5em}
\end{figure*}
The detailed steps of the solution are described in Algorithm~\ref{algo}.
\subsubsection{Algorithm complexity} The proposed algorithm solves the chance constrained optimization 
problem in~\eqref{eq_problem2} through a two-step decomposition. First, the integer variable $N_{t_{q'}}$ is enumerated over 
$\{1, \ldots, N_{\max}\}$, contributing a factor of 
$\mathcal{O}(N_{\max})$. Second, for each fixed $N_{t_{q'}}$, the reduced power allocation subproblem in~\eqref{eq_power_allocation} 
is solved via bisection over $[0, P_{q}^{\max}]$ to numerical tolerance $\varepsilon$, requiring $\mathcal{O}(\log(1/\varepsilon))$ 
iterations. At each bisection iteration, the KKT stationarity condition in~\eqref{eq_KKT} must be evaluated across all 
$q \in \{1, \ldots, m_{q,q'}\}$ power allocation threshold levels, where each evaluation involves computing the partial derivatives 
$\partial f(P_q)/\partial P_q$ and $\partial g(P_q)/\partial P_q$ from \eqref{eq_KKT} costing $\mathcal{O}(m_{q,q'})$ per iteration. Combining these three factors, the per-slot complexity of the proposed algorithm is $\mathcal{O}(N_{\max} \cdot m_{q,q'} \cdot \log(1/\varepsilon))$, and since the optimization is executed dynamically at each of the $T$ jitter state evolution steps in Algorithm~1, the total complexity is $\mathcal{O}(T \cdot N_{\max} \cdot m_{q,q'} 
\cdot \log(1/\varepsilon))$, where $T$, $N_{\max}$, $m_{q,q'}$, and $\varepsilon$ denote the number of transmission slots, the maximum antenna count, the total number of power allocation threshold levels, and the bisection tolerance, respectively. Also, in solving \eqref{eq_power_allocation}, $N_{\mathrm{max}}$ is set to 1. Similarly, with $P_q$ fixed, the optimal value of $N_{t_{q'}}$ can be obtained.

\vspace{-.9em}
\section{Simulation Results and Analysis}\vspace{-0.em}
This section details the resilience analysis and Monte Carlo evaluation of the V2V link across alarming, failure, and restoration phases of a single resilience cycle. Random processes like interference growth, distance variations, and delay jitter dynamics are simulated in MATLAB, with key performance metrics plotted and analyzed at each phase. Simulations use the parameters provided in Table I and averages from an ensemble of independent trials. All the simulated plots are obtained for $10^6$ Monte-Carlo trials.

\vspace{-1.2em}
\subsection{System performance during alarming stage}
During the alarming stage, the V2V link begins to experience transmission-delay jitter driven by the stochastic growth of both inter-vehicular distance and interference. As these deterioration processes intensify, ARER rises, while $\mathcal{C}(t)$ correspondingly declines. Figures~3 and 4 show that at approximately the $32-34$ and $50-55$ transmission slots, $\mathcal{C}(t)$ and $\mathcal{D}(t)$ intersect for interference and V2V distance variation, respectively. At these instants, the limit-state indicator, $\mathcal{G}(t)$, crosses zero, signaling a critical jitter condition. Beyond these, $\mathcal{G}(t)$ becomes increasingly negative, which implies that $\mathcal{D}$ exceeds the available capacity $\mathcal{C}$ and the considered V2V system is not capable of tolerating the stressor-induced degradation. 

Figures~5 and 6 illustrate how $\mathcal{C}(t)$ evolves during the alarming phase under different values of active transmit antennas, $N_t$, which reveals the influence of spatial diversity on jitter resilience. In Figure~5, where distance-induced fading dominates, increasing the number of active transmit antennas significantly slows the decline in $\mathcal{C}(t)$. This result highlights the effectiveness of spatial diversity in mitigating the impact of stochastic distance variations. It can be observed from Fig.~5 that, around the $60^{\mathrm{th}}$ time slot and in the presence of random V2V distance variations, $\mathcal{C}(t)$ of the considered V2V system is approximately $0.15$, $0.35$, $0.43$, and $0.65$ for $N_t=1, 10, 20,~\text{and}~30$, respectively. Therefore, larger antenna arrays provide increased resistance (with improved $\mathcal{E}_r$) to mobility-induced V2V jitter degradation, thereby enhancing the system's resilience to the adverse effects of growing inter-vehicular separation on jitter intolerance performance. In contrast, Figure~6 reveals an opposite trend under interference-limited stress conditions. In this case, increasing $N_t$ accelerates performance degradation. For example, at $40^{\mathrm{th}}$ transmission slot and depending upon the interference intensity, $\mathcal{C}=0.22$ when $N_t=30$, while $\mathcal{C}=0.49$ when $N_t=1$. Because the additional transmit power from more active antennas amplifies interference more rapidly than the achievable diversity gain can compensate, thereby diminishing the system's jitter tolerance. It can be inferred from \eqref{del_jit_dist} and \eqref{del_jit_int} that when a single stressor dominates, either distance variation $((1-p_I)p_d)$ or interference variation 
$((1-p_d)p_I)$, an increase in the number of transmit antennas attenuates the sensitivity of the jitter state to distance-induced fluctuations, $\frac{\delta \tau_{\mathrm{tr}}(t)}{\delta \varsigma_d}$, while enhancing its sensitivity to interference-induced changes, $\frac{\delta \tau_{\mathrm{tr}}(t)}{\delta \zeta_I}$, respectively. Thereby, this counterintuitive outcome also highlights that greater resource availability does not necessarily yield better performance and can even undermine resilience in interference-dominated environments. Figures~5 and 6 show two distinct degradation patterns. Distance-driven degradation in Figure~5 is gradual and manageable through diversity, whereas interference-driven degradation in Figure~6 leads to rapid resilience loss under early resource overuse. This indicates that interference is the main factor limiting V2V jitter performance in the considered multi-link setup. Hence, identifying the dominant stressor is essential, as antenna diversity can improve performance under mobility-induced fading but degrade it under interference-dominated conditions. 

 \vspace{-1.em}
\subsection{Restoration stage}
When $\mathcal{G}$ falls below zero, the V2V link initiates resource allocation to regain $\mathcal{C}(t)$. In our simulations, two corrective actions are implemented: (i) adaptive transmit power control and (ii) activation of multiple transmit antennas. Figures~7 and 8 demonstrate that either strategy is able to reverse the downward capacity trend caused by distance-driven fading or escalating interference. After allocating the resources adaptively, $\mathcal{G}$ improves steadily and eventually reaches the desired upper threshold of~$1$, confirming that both power adaptation and antenna diversity can fully restore $\mathcal{C}(t)$ once a failure event has been detected. It can be seen from Figs.~7 and 8 that, during the restoration phase, adaptive resource allocation enhances $\mathcal{C}(t)$ in several stages and eventually drives it to its maximum achievable value of $1$. This behavior arises because the intensity of both stressors continues to grow randomly even after failure, so the ongoing deterioration shapes the stochastic evolution of the system's jitter‑intolerance state. 

From Figs.~7 and 8, it can be observed that during the alarming phase, when the considered V2V link operates without any adaptive resource-allocation mechanism and relies only on fixed resource parameters (e.g., transmit power or number of wireless channels), the impact of stressors continues to grow. Consequently, $\mathcal{D}(t)$ increases from around $0\%$ to $50\%$ under the considered worst-case limit-state boundary condition $(\mathcal{G}=0)$, and the accumulated jitter load perturbs $\mathcal{C}(t)$ for approximately $50$ transmission slots. In contrast, when adaptive resource allocation is implemented with the knowledge of stressor-driven $\mathcal{P}_{JI}$ variation, it is shown in Fig.~7 that $\mathcal{D}(t)$ decreases from $50\%$ to $0\%$ and $\mathcal{C}(t)$ can be restored in approximately $5$ transmission slots for the considered set-up. Moreover, adaptive resource allocation can significantly improve ARER performance by reversing the trend of $\mathcal{C}(t)$ degradation. As it is established in \eqref{ARER}, the positive values of ARER indicate the increasing jitter performance loss $(\mathcal{P}_{JI}^+(t)>\mathcal{P}_{JI}^-(t))$. While the adaptive resourse allocation can make ARER negative $(\mathcal{P}_{JI}^+(t)<\mathcal{P}_{JI}^-(t))$, changing the slope from approximately $0.00956$ to $-0.03026$ per transmission slot. This improvement restores stable jitter performance in the considered V2V system, as shown in Fig.~\ref{fig9}. Consequently, the ARER gain during the recovery phase with adaptive resource allocation is approximately $3.17$ times higher than that observed during the alarming phase under constant resource allocation.

\setcounter{figure}{8}
\begin{figure}[t]
\centerline{\psfig{file=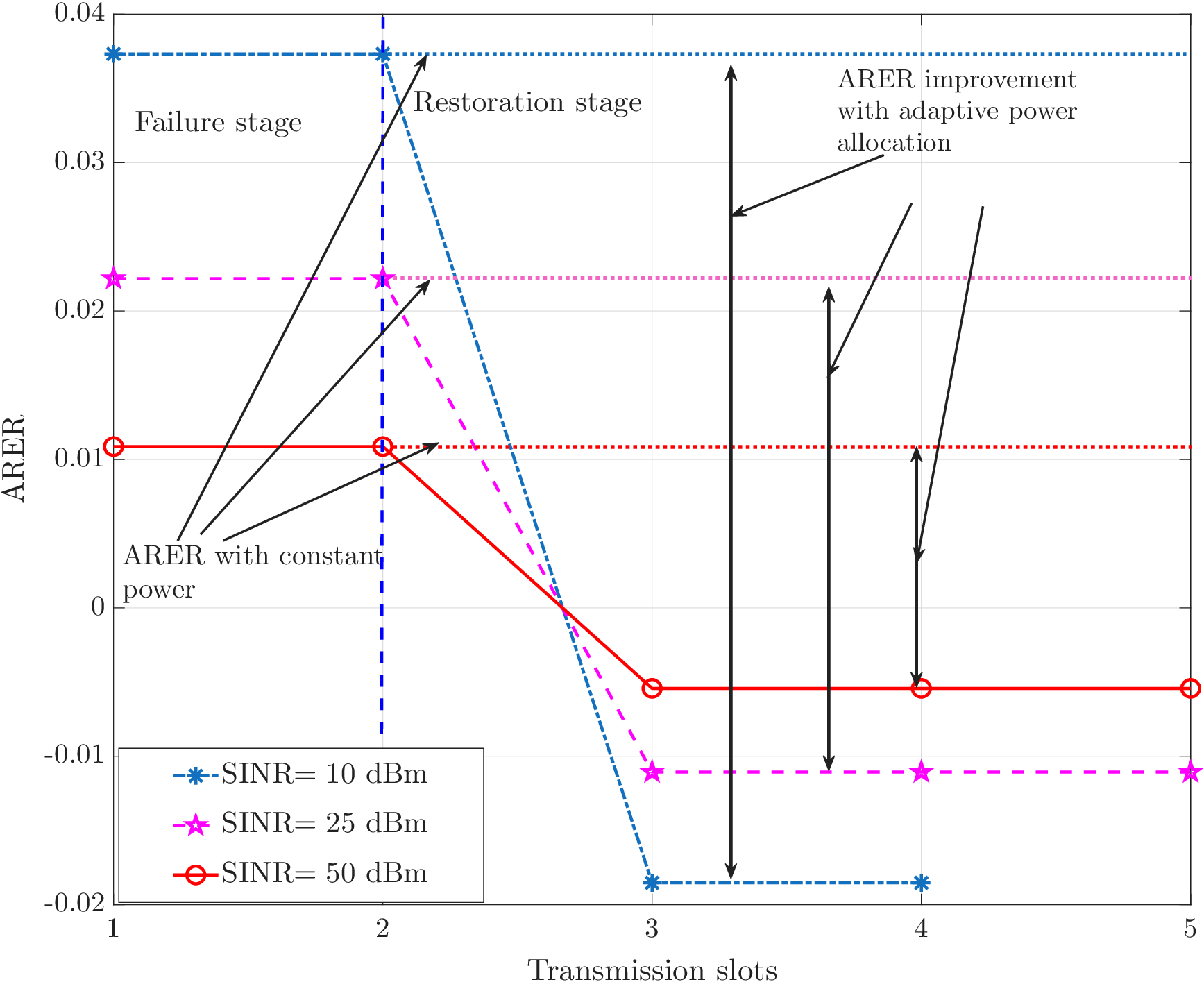,width=2.6 in, height = 2.1 in }}	\vspace{-0.5em}
\caption{\textcolor{black}{{\small $\mathcal{E}_r(t)$ with constant power and adaptive power allocation.}}} \label{fig9} \vspace{-1.7em}
\end{figure}

\vspace{-1em}
\section{Conclusion}
In this paper, we have studied how mobility and interference-driven deterioration processes affect the transmission delay jitter profile in a V2V communication link. We developed stochastic models for the degradation mechanisms and analyzed the time evolution of the V2V link jitter state. To characterize the joint effect of deterioration intensity and jitter, we introduced a limit-state function, $\mathcal{G}$, that tracks $\mathcal{C}$ over time. Monte-Carlo simulations confirm the accuracy of the analytical framework and show that adaptive transmit-power control and implementing multiple transmit links can improve $\mathcal{C}(t)$ and restore $\mathcal{G}(t)$ to its safe operating range after the system enters the critical performance region, thereby significantly improving the considered system's jitter intolerance state. In future work, we will investigate recovery rate analysis by integrating adaptive jitter mitigation techniques with hybrid resource allocation strategies, particularly under varying environmental conditions where scattering and mobility introduce additional challenges to the V2V wireless channel.

\vspace{-1em}
\appendices       
\section{Laplace approximations}
\noindent From \eqref{jit_intol_erf}, $y_1$ can be written as
$
y_1=\frac{p_I p_d}{2}(1+\text{erf}(x_1))=g(x_1).$
Lets consider $x_{10}$ is the mode of $x_1$, can be theoretically expressed from \eqref{mean_del_jit_next} as
$x_{10} = \mathbb{E}[\tau_{\mathrm{tr}}(t^+)] = \bigg(\frac{\partial \tau_{\mathrm{tr}}(t)}{\partial \zeta_I} \mu_{\zeta_I} \hspace{-0.3em}+\hspace{-0.2em} \frac{\partial \tau_{\mathrm{tr}}(t)}{\partial \varsigma_d} \mu_{\varsigma_d}\hspace{-0.3em}\bigg).$ Therefore, near the value $0$, $y_{10}=g(x_{10})$ and locally $x_1\approx N(x_{10},\sigma_{x_1}^2)$, where $\sigma_{x_1}^2=\bigg(\frac{\partial \tau_{\mathrm{tr}}(t)}{\partial \zeta_I}\bigg)^2 \sigma_{\zeta_I}^2 + \bigg(\frac{\partial \tau_{\mathrm{tr}}(t)}{\partial \varsigma_d}\bigg)^2 \sigma_{\varsigma_d}^2$ (from \eqref{var_del_jit_next}). Now considering $h(x_1)=\log f_{X_1}(x_1)$ and applying Taylor expansion around $x_{10}$, we get
$h(x_1)=h(x_{10})+h'(x_{10})(x_1-x_{10})+\frac{h''(x_{10})}{2!} (x_1-x_{10})^2.$
Since $x_{10}$ is the mode of $x_1$ (local maxima), $h'(x_{10})=0$ and $h''(x_{10})=-ve=-k$, where $k$ is a constant parameter. Thereby, the PDF of $x_1$ is deduced as: \vspace{-0.5em}
\begin{IEEEeqnarray}{rCl}
f_{X_1}(x_1) \approx \sqrt{\frac{k}{2 \pi}} e^{\frac{k}{2}(x_1-x_{10})^2} \approx N(x_{10}, \sigma_{x1}^2=1/k). \vspace{-0.7em}
\end{IEEEeqnarray}
Next, we can express $y_1$ utilizing $g'(x_1)=\frac{p_I p_d}{\sqrt{\pi}}e^{-x_1^2}$ and Taylor series expansion as
$
y_1=y_{10}+g'(x_{10})(x_1-x_{10}).$
After some computation, from (56) we get $\frac{dx_1}{dy_1}=\frac{1}{g'(x_{10})}$. Therefore, the PDF of $y_1$ can be computed as:\vspace{-0.5em}
\begin{IEEEeqnarray}{rCl}
f_{Y_1}(y_1) \approx \sqrt{\frac{k}{2 \pi}} e^{\frac{k}{2}(x_1-x_{10})^2} \bigg|\frac{1}{g'(x_{10})}\bigg|=\frac{1}{\sqrt{2\pi\sigma_{y_1}^2}}e^{\frac{-(y_1-y_{10})^2}{2\sigma_{y_1}^2}}, \nonumber\\& \vspace{-0.8em}
\end{IEEEeqnarray}
where $\sigma_{y_1}^2=\frac{(p_I p_d)^2 e^{-2x_{10}^2}}{k \pi}$ and $y_{10}=g(x_{10})=\frac{p_I p_d}{2}(1+\text{erf}(x_{10}))$. Similarly, the PDFs of $y_2$ and $y_3$ can also be obtained.

\begingroup
\renewcommand{\baselinestretch}{0.88}\normalsize
\vspace{-1.em}\bibliographystyle{IEEEtran}
\bibliography{bibo}
\endgroup

\end{document}